\documentclass[prl,twocolumn,superscriptaddress,floatfix,noshowpacs,10pt,longbibliography]{revtex4-1}%

\usepackage{graphicx,bm,times}
\usepackage{amsmath}
\usepackage{amsfonts}
\usepackage{amssymb}

\usepackage{bm}
\usepackage{color}
\usepackage{times}

\usepackage[%
colorlinks=true,
urlcolor=blue,
linkcolor=blue,
citecolor=blue
]{hyperref}

\begin{document}

\title{Electronic nematic states tuned by isoelectronic substitution in bulk FeSe$_{1-x}$S$_x$}

\author{Amalia I. Coldea}
\email[corresponding author: ]{amalia.coldea@physics.ox.ac.uk}
\affiliation{Clarendon Laboratory, Department of Physics,
University of Oxford, Parks Road, Oxford OX1 3PU, UK}

\begin{abstract}
Isoelectronic substitution is an ideal tuning parameter to alter electronic states and correlations in iron-based superconductors.
As this substitution takes place outside the conducting Fe planes,
the electronic behaviour is less affected by the impurity scattering experimentally
and relevant key electronic parameters can be accessed.
In this short review, I present the experimental progress made in understanding the electronic behaviour of the
  nematic electronic superconductors, FeSe$_{1-x}$S$_x$.
 A direct signature of the nematic electronic state is in-plane anisotropic distortion of the Fermi surface
triggered by orbital ordering effects and electronic interactions that result in
multi-band shifts detected by ARPES.
Upon sulphur substitution,
the electronic correlations and the Fermi velocities decrease in the tetragonal phase.
Quantum oscillations are observed for the whole series in ultra-high magnetic fields and show a complex spectra due to the presence
of many small orbits.
Effective masses associated to the largest orbit display non-divergent behaviour at the nematic end point
 ($x \sim 0.175(5)$), as opposed to critical spin-fluctuations in other iron pnictides.
Magnetotransport behaviour has a strong deviation from the Fermi liquid behaviour
and  linear $T$ resistivity is detected at low temperatures inside the nematic phase,
 where scattering from low energy spin-fluctuations are likely to be present.
The superconductivity is not enhanced in FeSe$_{1-x}$S$_x$
 and there are no divergent electronic correlations at the nematic end point.
 These manifestations indicate a strong coupling with the lattice in FeSe$_{1-x}$S$_x$
 and a pairing mechanism likely promoted by spin fluctuations.
\end{abstract}
\date{\today}
\maketitle

\section{\bf Introduction}

Iron-based superconductors
offer a unique playground to understand unconventional superconductivity and
explore the normal competing electronic phases, such as nematic electronic phases and spin-density wave phases.
Often the nematic and spin-density phases
neighbour each other in the phase diagrams of iron-based superconductors,
making it difficult to assess whether the spin or nematic fluctuations are the most
relevant for stabilizing superconductivity \cite{Fernandes2014}.

The nematic electronic state of iron-based superconductors
breaks the rotational symmetry of the tetragonal Fe plane lattice
 from four-fold symmetric ($C_4$)  down to two-fold symmetric ($C_2$) \cite{Fradkin2010}.
  This symmetry breaking is expected
  to have a number of consequences
 on the electronic properties leading to
 a series of effects involving anisotropic single-particle properties,
showing a distorted Fermi surface (that can be triggered like a Pomeranchuk instability
in the presence of interactions \cite{Pomeranchuk1959}),
  anisotropic spin-fluctuation spectra, and
 anisotropic transport properties that can lead to non-Fermi-liquid behaviour \cite{Vojta2009}.
 Theoretically, in the proximity to a nematic quantum critical point,
the nematic fluctuations with
wave-vector $q = 0$ can enhance the
critical temperature by pairing through the
exchange of nematic fluctuations in all symmetry channels \cite{Lederer2017,Lederer2015}.
In real systems, the nematic electronic phase is intimately
coupled with the lattice. This coupling has significant consequences
on the observed response, such as the presence
of the tetragonal-to-orthorhombic structural transition
at the same temperature where the nematic electronic order develops.
This finite coupling of the electronic system with the lattice is expected
to alter  the response of the nematic critical fluctuations on superconductivity and
 the non-Fermi liquid power-law dependencies in transport \cite{Paul2017,Labat2017,Carvalho2019,Wang2019}.

Isoelectronic substitution is a clean and efficient way to tune phase diagrams of iron-based superconductors, by
gently suppressing the relevant electronic interactions and competing electronic phases with superconductivity, and to access quantum critical points \cite{Abrahams2011}.
A unique system, the chalcogenides FeSe$_{1-x}$S$_x$,
provides an essential  route to investigate the interplay between nematicity and superconductivity,
in the absence of long-range magnetism.
Furthermore, the isoelectronic substitution can access the
experimental manifestations around a putative nematic critical point,
undisturbed by the presence of a magnetic critical point,
as found in other systems, like BaFe$_2$(As$_{1-x}$P$_x$)$_2$ \cite{Kasahara2010}.
The parent compound of this family, FeSe, displays
a nematic electronic phase and a tetragonal-to-orthorhombic transition below 90~K \cite{Bohmer2013}
and no long range magnetic order was detected despite a rich spectrum of low and high-energy spin fluctuations \cite{Chen2019,Wang2016b,Kreisel2020}.
The bulk superconductivity of FeSe has a relatively low critical temperature close to 9~K
but it can be enhanced towards 40~K by applied external pressure  \cite{Medvedev2009,Sun2016pressure}.
The nematic phase of FeSe is also suppressed at low pressures
\cite{Medvedev2009,Terashima2015,Sun2016pressure,Kothapalli2016}
before a new magnetic state is stabilized at high pressures \cite{Bendele2012,Kothapalli2016},
that competes with the high-$T_c$ phase  \cite{Sun2016pressure}.
Besides applied pressure, the bulk superconductivity of FeSe can be enhanced towards 40~K
via the intercalation between the van der Waals layers of a molecular spacer \cite{Burrard2013},
and by gating of thin flakes \cite{Lei2016}.
In a monolayer on FeSe, on a suitable substrate, the transition temperatures reach
record values towards  65~K; a strong interfacial electron-phonon coupling and a charge transfer
through the interface is proposed as a source for this two-dimensional high-$T_c$ superconductivity
 \cite{Huang2017,Rebec2017}. This effect is surprisingly absent in a monolayer of FeS \cite{Shigekawa2019}
 and in the absence of substrate in thin flakes of FeSe \cite{Farrar2020}.

\begin{figure*}[htbp]
	\centering
\includegraphics[trim={0cm 0cm 0cm 0cm},width=1\linewidth,clip=true]{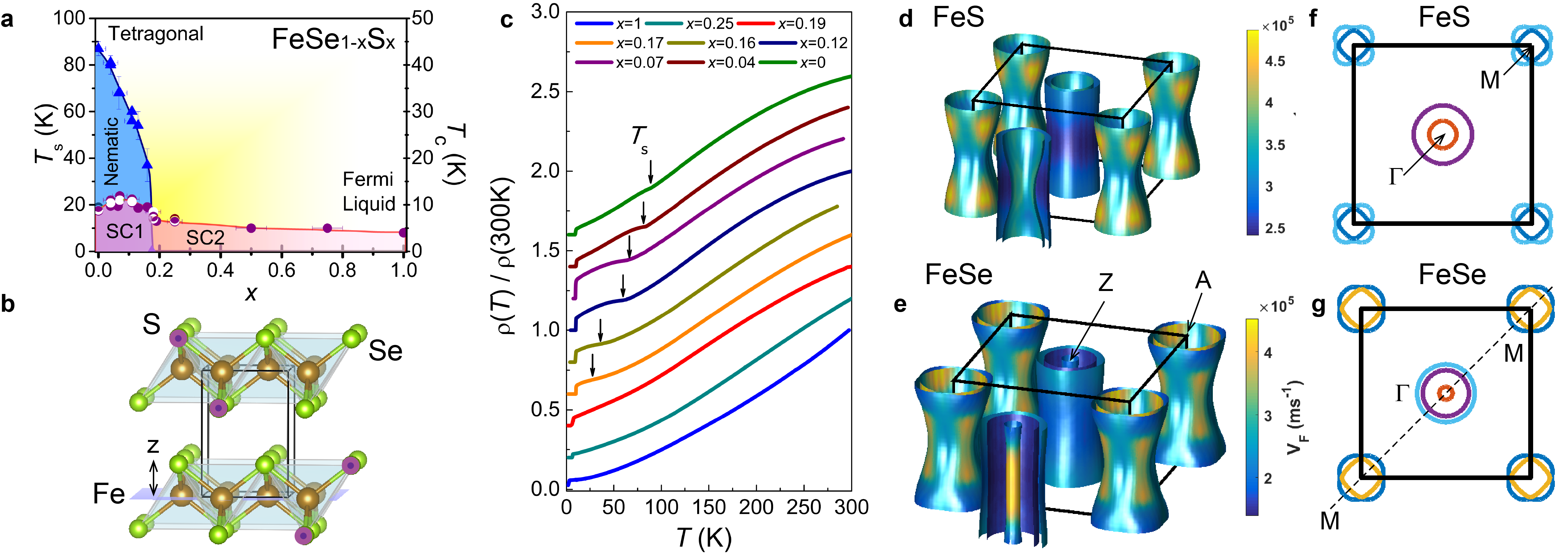}
	\caption{\textbf{Phase diagram of FeSe$_{1-x}$S$_x$.}
 a) Phase diagram of FeSe$_{1-x}$S$_x$ based on Ref.~\cite{Coldea2019}.
  c) The resistivity versus temperature of FeSe$_{1-x}$S$_x$ normalized at room temperature, after Ref.~\cite{Coldea2019,Bristow2020,Reiss2017}.
The data are shifted vertically and the arrows indicate the position of the nematic transition that coincides
with the tetragonal to orthorhombic transition at $T_{s}$. This transition is better visualized using the first derivative of the resistivity, as discussed in Ref.~\cite{Coldea2019,Bristow2020,Sato2018}.
 b) The tetragonal unit cell of FeSe$_{1-x}$S$_x$ (solid lines).
FeSe$_{1-x}$S$_x$ crystalizes in the $P4/nmm$ space group (No.129)
 with atoms Fe: 2a (3/4,1/4,0) and Se: 2c (1/4,1/4,$z_{Se}$).
 The position of the chalcogen above the Fe plane is indicated by the parameter $z$ that affects significantly the band structure.
Calculated Fermi surface using density functional theory (DFT using GGA approximation
 and spin-orbit coupling)  of FeS in d) and FeSe in e) using experimental lattice parameters at room temperature  for FeSe \cite{McQueen2009}  and
 FeS \cite{Lai2015,Guo2019}).
 The Fermi surfaces are coloured using the Fermi velocities indicated by the corresponding colour bar.
 The middle hole band with lower velocity in FeSe has a dominant $d_{xy}$ character whereas
the other hole and electron bands have mixed $d_{xz/yz}$ character.
The high-symmetry cut of the Fermi surface in the $\Gamma$-M plane ($k_z$=0)
for FeS in e) and FeSe in f). The dashed
line indicates the direction of the high symmetry cuts in ARPES measurements.}
	\label{fig_phase_diagram}
\end{figure*}

The isoelectronic substitution of the FeSe family can be achieved, by
replacing elements with a similar number of electrons outside the Fe planes,
using sulphur or tellurium ions on selenium ions sites.
The availability of single crystals of these materials have allowed intense interest and study of
their physical properties, summarized in recent reviews  in Refs.~\onlinecite{Coldea2017,Bohmer2017,Kreisel2020,Shibauchi2020}.
Furthermore, by combining physical and chemical pressures, the relative position of the nematic electronic phase
in relation to the spin-density wave phase can be varied and thus
the influence of two competing electronic phases on superconductivity can be disentangled \cite{Matsuura2017,Reiss2020}.
The scope of this review is to summarize the recent experimental efforts in understanding the electronic behaviour
of FeSe$_{1-x}$S$_x$ that can provide a unique insight into the
role played by the nematicity, Fermi surfaces, proximity to a putative nematic critical point
and electronic correlations in relation to superconductivity
in the absence of any long-range magnetic order.

\section{\bf Phase diagram of FeSe$_{1-x}$S$_x$}

Figure~\ref{fig_phase_diagram}a shows the phase diagram of  FeSe$_{1-x}$S$_x$
as a function of isoelectronic substitution with sulphur obtained from transport measurements.
The isoelectronic substitution  achieved by replacing selenium ions
for sulfur ions  outside the Fe plane causes an positive internal chemical pressure
as these ions have different ionic radii (S$^{2-}$ is 1.70~\AA~compared with 1.87~\AA~for Se$^{2-}$)
 \cite{Watson2015c,Mizuguchi2009}.
 The nematic electronic phase of FeSe,
  due to its finite coupling with the underlying lattice,
   triggers a structural transition from a
tetragonal to an orthorhombic phase at $T_{\rm s}$ \cite{Bohmer2013}.
This transition gives rise to a well-defined anomaly in the transport measurements (Figure~\ref{fig_phase_diagram}c)
that helps to build the nematic phase diagram
and to identify the expected position of each measured single crystal
inside the nematic phase, as shown in Figure~\ref{fig_phase_diagram}a.

The isoelectronic substitution with sulphur in FeSe leads to the efficient suppression of the
nematic electronic state, similarly to the effect of applied pressure \cite{Terashima2014} (Figure~\ref{fig_phase_diagram}a).
In contrast to applied pressure, the nematic phase can be completely suppressed with sulphur substitution
and no spin-density wave phase was detected for any available single crystals.
The lowest detected value of $T_{\rm s}$ is about 25~K, followed by
an abrupt drop at the nematic end point (NEP) which occurs close to $x \sim 0.175(5)$,
 \cite{Watson2015c,Hosoi2016,Reiss2017,Wiecki2018,Coldea2019}.
Thus, FeSe$_{1-x}$S$_x$ family is unique and permits the exploration of the
 nematic electronic phase transition in the vicinity
of a putative nematic critical point.

Inside the nematic phase, the superconducting transition temperature
displays a small dome reaching $T_{\rm c} \sim $~11~K close to $x \sim0.11 $,
varying from  8.7(3)~K for FeSe inside the nematic phase towards 6.5-5~K
just outside it \cite{Bristow2020Hc2,Reiss2017,Wiecki2018,Sato2018}.
For higher $x$ values inside the tetragonal phase, the superconductivity hardly changes reaching only 4.5(5)~K towards FeS
 \cite{Lai2015,Guo2019}.
 The suppression of the nematic phase transition in FeSe$_{1-x}$S$_x$
coincides with a decrease in the superconducting transition temperature $T_{\rm c}$  close to NEP.  STM studies have detected a rather abrupt  change in the superconducting  order parameter
 at the nematic phase boundaries,
 implying that  different types of pairing may be operational
  inside (SC1) and outside the nematic phase (SC2),
 as shown in Figure~\ref{fig_phase_diagram}a  \cite{Hanaguri2018,Sato2018}.
 Recently, it has been suggested theoretically that
 a  topological transition associated with the creation of a Bogoliubov Fermi surface
could occur as a function of $x$  in FeSe$_{1-x}$S$_x$ \cite{Setty2020,Setty2020PRB}.

In order to understand in depth the electronic properties
of FeSe$_{1-x}$S$_x$ a good knowledge of the
exact composition and the sulphur variation in each
batch is required.
This can be challenging for techniques, like neutron-diffraction
and muon spin rotation, that require a large mass of sample made of
hundreds of small single crystals \cite{Holenstein2019}.
 At room temperature
 FeSe$_{1-x}$S$_x$ crystalizes in the $P4/nmm$ space group (No.129),
 as shown in Figure~\ref{fig_phase_diagram}b.
The lattice parameters of FeSe are $a = 3.7651$ \AA, $c$ = 5.5178 \AA,   $z_{Se}$ = 0.2672 \cite{McQueen2009}
whereas FeS has a much smaller $c$ axis ($a = 3.6802 $ \AA, $c$ = 5.0307 \AA,   $z_{S}$ = 0.2523)  \cite{Lai2015,Guo2019}.
The lattice parameters measured by X-ray diffraction
for each crystal of FeSe$_{1-x}$S$_x$ can be used
to determine the composition of each sample.
Assuming the formation of a solid solution as a function of composition,
the values of the  lattice parameters at room temperature, $p$,
(that can be $a,b$ and $c$ or $z_{\rm Se/S}$)
for a certain composition $x$ can be estimated
using an empirical Vegard's law
$p_{x}$=$x~p_S$+(1-$x$) $p_{\rm Se}$.

\subsection{Single crystal growth}
Single crystals of FeSe$_{1-x}$S$_x$
are normally grown by the KCl/AlCl$_3$ chemical vapor transport method from the FeSe end towards
$x \sim 0.4$ \cite{Chareev2013,Bohmer2013,Hosoi2016,Wiecki2018,Watson2015c,Reiss2017,Bohmer2016g}.
The growth of higher concentrations and FeS was achieved
using a hydrothermal reaction of iron powder with sulfide solution, which in general is a more
invasive method and can lead to single crystals with higher concentration of impurities \cite{Lai2015,Guo2019,Reiss2017}.
Epitaxial thin films of FeSe$_{1-x}$S$_x$ with $\leq 0.43$
were grown via pulsed laser deposition \cite{Nabeshima2018}.
A potential  anomaly was observed in the resistivity
curves for films with large $x$,
suggested to be linked to a magnetic transition \cite{Nabeshima2018},
but these findings have not been yet confirmed in single crystals \cite{Wiecki2018,Nabeshima2018}.
The exact $x$ composition for samples in each  batch is normally checked using compositional analysis
using  energy-dispersive X-ray spectroscopy (EDX) or
electron-probe micro-analysis  (EPMA)~\cite{Chareev2013,Bohmer2013,Wiecki2018,Coldea2019,Nabeshima2018}.
The nominal composition, $x_{\rm nom}$, used during the
growth process  is  often smaller than the real composition $x$ (by about 80\%)  and
the higher the composition the larger degree of variation occurs
within the same batch \cite{Wiecki2018,Nabeshima2018}.
For example, the phase diagrams of FeSe$_{1-x}$S$_x$
reported in Refs.~\cite{Licciardello2019,Licciardello2019MR} uses the nominal values  $x_{\rm nom}$.
Thus, the linear resistivity in 35~T occurs {\it inside} the nematic phase, as the measured
zero resistivity shows an anomaly at  $T_{\rm s}\sim$ 51~K for a nominal composition $x_{\rm nom}\sim 0.16$
that would correspond around $x \sim 0.13$ \cite{Bristow2020}.
The residual resistivity ratio, defined as the ratio between room temperature resistivity
and the resistivity at the onset of superconductivity,
varies between 15-44 \cite{Coldea2019,Bristow2020}
and it is often used as a proxy to assess the quality of each single crystal.
In high magnetic field, quantum oscillations were observed
for all composition of FeSe$_{1-x}$S$_x$ reflecting their high quality
with large mean free path (up to $\sim $ 350~\AA) \cite{Coldea2019,Reiss2020}.
For higher $x$ composition,
the mean free path decreases slightly and
new hexagonal phases could be stabilized \cite{Coldea2019}.
The superconductivity of
Fe$_{1+\delta}$Se can also be destroyed by very
small changes in its stoichiometry \cite{McQueen2009}.

\section{\bf Electronic Structure of FeSe$_{1-x}$S$_x$}

The main features of the electronic structure  of FeSe$_{1-x}$S$_x$
can be understood by considering the two-dimensional square lattice of Fe ions,
separated by Se/S atoms residing above and below the Fe layer, as shown in Figure~\ref{fig_phase_diagram}b.
Due to the strong bonding between the Fe-Fe and Fe-(Se/S) sites,
an Fe atom can be placed inside the centre of an almost perfect tetrahedron of Fe(Se/S)$_4$
that determines  the electronic properties of these materials.
Band structure calculations show that the Fermi surface of FeSe$_{1-x}$S$_x$
consists of well-separated hole pockets at the center of the Brillouin zone and electron pockets at the zone corners
that form quasi-two dimensional Fermi surfaces, as shown in Figure~\ref{fig_phase_diagram}d and e.
The position of the chalcogen ion in relation to the Fe plane, $z$, affects significantly the predicted
number and the orbital character of the hole bands,
FeSe having an additional middle hole band with $d_{xy}$ character which is pushed below the Fermi level in FeS (Figure~\ref{fig_phase_diagram}d and e).
There are two predicted cylindrical electron bands which hardly change in shape across this series, similar to the
isoelectronic series BaFe$_2$(As$_{1-x}$P$_x$)$_2$ \cite{Shishido2010}.
The positive chemical pressure in FeSe$_{1-x}$S$_x$ results in a lattice
contraction and the reduction of the $c$ axis \cite{Mizuguchi2009}
and it would bring the Fe(Se/S) layers closer together,
 increasing the bandwidth and potentially
 leading to the suppression of the electronic correlations \cite{Reiss2017,Miao2017}.
As discussed below, DFT calculations provide essential guide to understand
the origin of the observed Fermi surfaces of FeSe$_{1-x}$S$_x$, but the size
are smaller, the number of hole bands is reduced compared with calculations and the
$k_z$ dependence is changed.

\begin{figure*}[htbp]
\centering
       	\includegraphics[trim={0cm 0cm 0cm 0cm}, width=1\linewidth,clip=true]{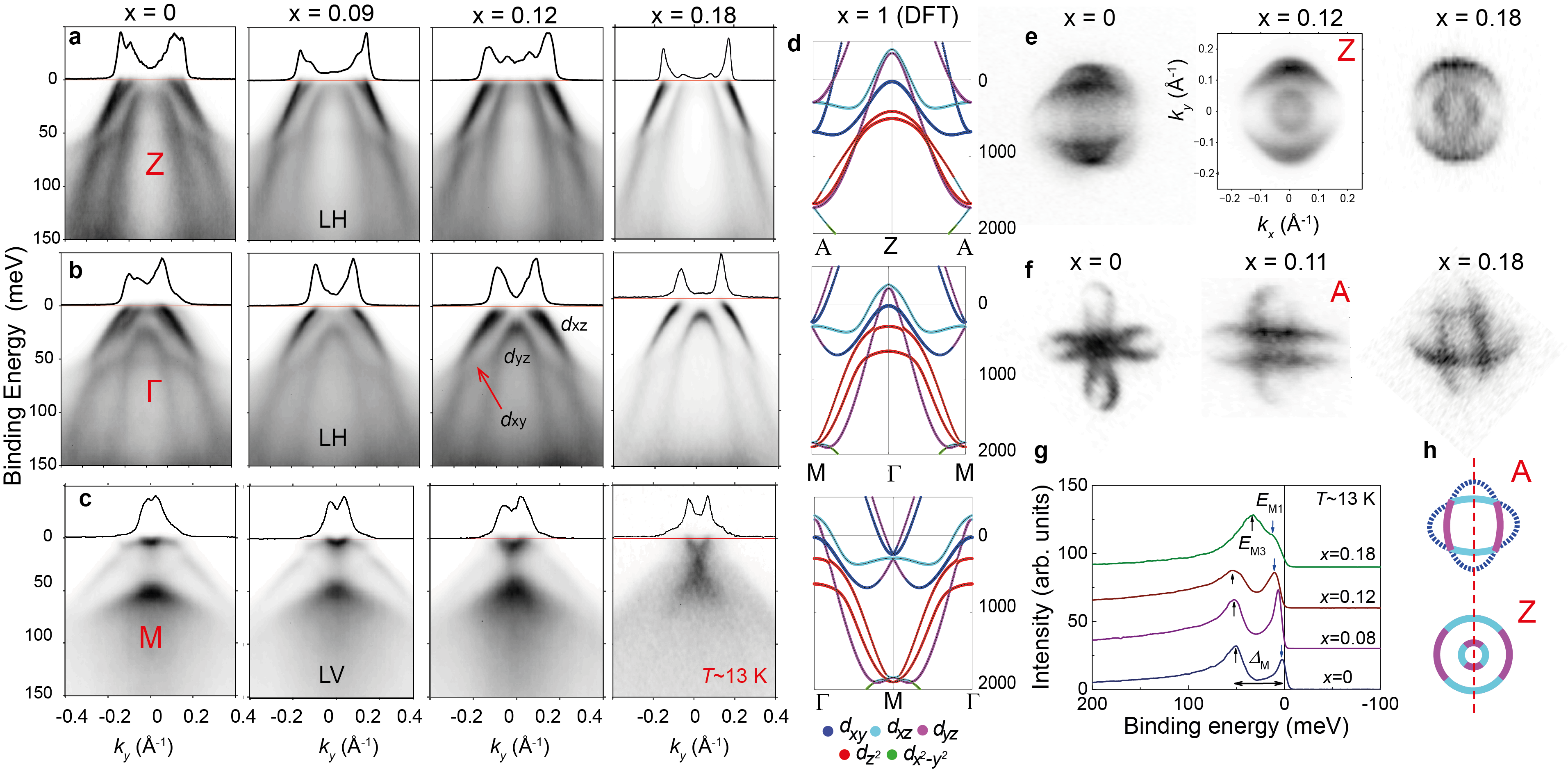}
       	 \caption{\textbf{Low-temperature ARPES data for FeSe$_{1-x}$S$_x$.}
Band dispersion around high symmetry points, a) $\Gamma$, b) $Z$
and c) M collected using horizontally and vertically
linearly polarized synchrotron light (LH and LV) at low temperatures at 13~K
for  different $x$ compositions (after Refs.~\onlinecite{Watson2015c,Reiss2017,Coldea2017}).
Momentum dependent cuts (MDCs) are shown above each dispersion which allows to extract
the $k_{\rm F}$ values at the Fermi level (the instrumental resolution is $\sim 3$~meV).
The top of the parabolic band dispersion
for the outer hole band at the $\Gamma$ point for $x$=0.18 is getting closer to the Fermi level
($\sim 5$~meV) compared with FeSe ($\sim 17$meV).
d) DFT calculations high symmetry cuts for FeS for the same symmetry points
like the ARPES data with energies in meV.
These dispersions can be compared with experiments
to extract the renormalization factor for each band.
The colours indicate the different orbital character.
The Fermi surface map for the hole bands at the Z point in e) and for the electron bands at the A point
in f), after Refs.~\onlinecite{Watson2015c,Reiss2017,Watson2016,Coldea2017}.
Inside the nematic phase, the existence of twining below $T_s$ can cause superposition
of the signal from two domains, rotated by 90$^{\circ}$.
The orbital character of the map varies around the Fermi surface
as depicted for the tetragonal case in h).
The dashed line indicates the cuts in a)-c) and the orientation
of the maps in e) and f).
g) The energy dependent cuts (EDCs)
centered at the high symmetry M point for different composition. The energy separation between the two most intense peaks
is defined by $\Delta_{M}$, a proxy for the orbital ordering effect.
The positions of the high symmetry M1 and M3
points for $x=0.18$ are indicated by arrows. h) Schematic of the tetragonal Fermi surfaces
in the Z-A plane containing different orbital characters.
The the dashed line is a cut along the diagonal of the unit cell used in
the ARPES measurements in panels a-c. }
 \label{Fig_ARPES}
\end{figure*}

\section{\bf ARPES studies of FeSe$_{1-x}$S$_x$}

ARPES is highly suited for the exploration of FeSe$_{1-x}$S$_x$ as these systems can be easily cleaved
in-situ due to weak van der Waals bonds between the FeSe layers
which also enable the development of devices of
 two-dimensional superconductors  by mechanical exfoliation \cite{Lei2016,Farrar2020}.
Furthermore, ARPES studies can evaluate the role of orbital character on the nematic electronic states,
as the matrix element effects affect the intensity of different bands with different orbital character.
In certain conditions, ARPES spectra of iron-based superconductors does not show certain branches
due to the underlying symmetry, in particular for the electron bands \cite{Sobota2020,Brouet2012,Moreschini2014},
as the intensity depends strongly on the polarisation of the incident beam
 as well as  the incident photon energy.
 A representation of the orbital character of different pockets at high symmetry points is shown
in Fig.~\ref{Fig_ARPES}h.

Extensive experimental ARPES studies on FeSe
found that system has many relevant electronic ingredients for a multiband system  \cite{Coldea2017,Watson2015a,Watson2017,Shimojima2014,Watson2017strain,Yi2019,Sobota2020}.
The experimental Fermi surface of FeSe is unusually small having
two electron pockets and a single hole pocket (instead of 3),
a factor 5 smaller than that  predictions of the band-structure calculations (Fig.~\ref{fig_phase_diagram}e).
Such a small Fermi surface could be sensitive to topological changes
in magnetic fields or under applied strain.
To bring the DFT calculated Fermi surfaces in agreement with experiments,
band shifts need to be applied in opposite
direction for hole and electrons of more than 200~meV for FeSe \cite{Watson2015a}
and less than 100~meV for FeS \cite{Terashima2019}.
Band shifts also occur at high temperatures inside the tetragonal phase of FeSe
and these effects are caused by higher energy interactions \cite{Kushnirenko2017}
as well as  the changes in the chemical potential  \cite{Rhodes2017},
as found in many iron-based superconductors \cite{Brouet2013}.
Furthermore, like many other iron chalcogenides, FeSe exhibits
strong orbitally-dependent electronic correlations
due to the larger band renormalization factor $\sim 7-9$
of the $d_{xy}$ band compared with $\sim 3-4$ for the $d_{xz/yz}$ band \cite{Maletz2014,Watson2015a}.
These values are obtained by comparing the experimental band dispersion to those
from  DFT calculations in the tetragonal phase, as shown in Figure~\ref{Fig_ARPES}
\cite{Maletz2014,Watson2015a}.
At high binding energies, ARPES spectra detected Hubbard-like bands
suggesting the existence of incoherent many-body excitations originating from
 Fe 3$d$ states, in addition to the renormalized quasiparticle bands near the Fermi level \cite{Watson2017,Evtushinsky2016}.
  Many high energy features of the observed ARPES data can be accounted for
  by considering  the strong local Coulomb interactions on the spectral function via dynamical mean-field theory,
  including the formation of a Hubbard-like band \cite{Watson2017,Evtushinsky2016}.
Another inherent challenge for ARPES studies inside the nematic phase
is the likely presence of sample twinning (rotated by $90^{\circ}$), by cooling thorough the structural transition,
and a lot of recent effort has been dedicated to address this issue by applying strain to FeSe
\cite{Watson2017strain,Yi2019,Cai2020}.

\subsection{\bf Hole pockets of FeSe$_{1-x}$S$_x$}
\begin{figure*}[htbp]
	\centering
    	    \includegraphics[trim={0cm 0cm 0cm 0cm}, width=1\linewidth,clip=true]{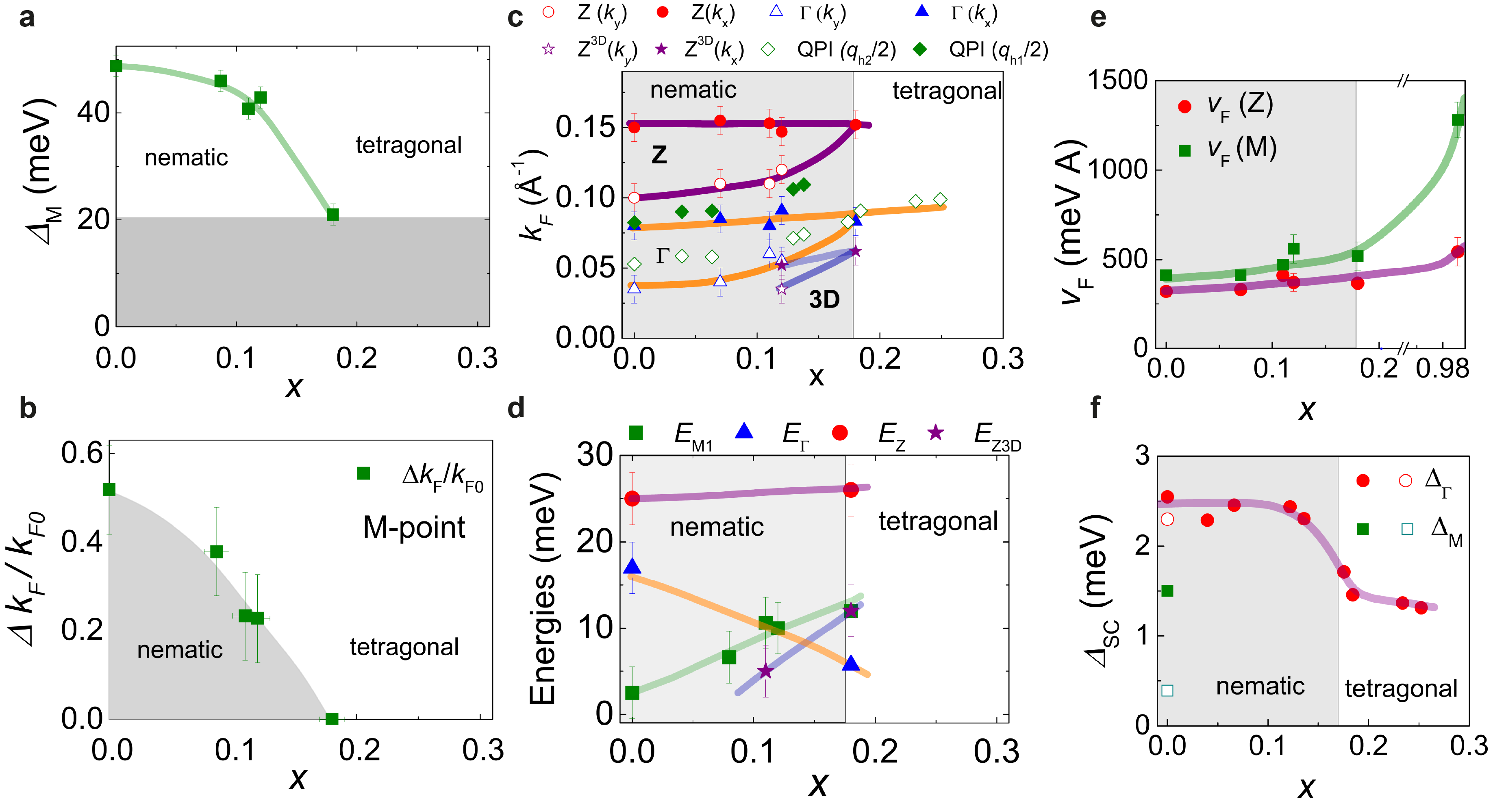}
	\caption{{\bf Low temperature parameters extracted from APRES data of FeSe$_{1-x}$S$_x$.}
 a) The variation of the low temperature splitting parameter, $\Delta_M$, extracted from EDCs at the M point, with $x$ substitution,
 shown in Fig.~\ref{Fig_ARPES}g.   A splitting of 20~meV is already present for the tetragonal $x$=0.18 suggesting the effect of nematicity is less than $\Delta_M$ for $x$=0 ~\cite{Reiss2017,Watson2015c}.
    b) The relative distortion of the electron Fermi surface at the M point, $\Delta{}k_F/k_{F0}$, at $T \sim $13~K
    in relation with the tetragonal phase $k_{F0}$, as a function of $x$ substitution (after Ref.~\onlinecite{Watson2015c}).
    c) The variation of the Fermi $k_F$ vector with $x$ for the hole pockets
   extracted from the MDC cuts and assuming that the data arises from two superimposed ellipses.
    The value of the $q/2$ scattering vectors from
    QPI are compared with those from ARPES data, after Ref.~\cite{Hanaguri2018}.
    d) The variation of the energy of the top or bottom of the bands at different high-symmetry points.
    e) The Fermi velocities extracted form the slope of the energy dispersion as a function of momentum
    for the hole and the electron bands, after Ref.~\cite{Reiss2017,Watson2015c}.
    f) The $x$ variation of the superconducting gap, $|\Delta_{SC}|$, extracted for the hole band for FeSe$_{1-x}$S$_x$
    and the electron band for FeSe, after Ref.~\cite{Hanaguri2018,Sprau2016}.
    Another potential small gap of 0.39~meV (open square) was invoked from specific heat data for FeSe \cite{Sun2017}.  }
	\label{Fig_ARPES_paramaters}
\end{figure*}

The evolution of the hole pockets of FeSe$_{1-x}$S$_x$ with $x$ substitution for
the two high symmetry points Z (at the top of the Brillouin zone) and $\Gamma$ (at the centre of the Brillouin zone)
at low temperatures is shown in Fig.~\ref{Fig_ARPES}(a) and (b), respectively.
The observed energy dispersions of FeSe$_{1-x}$S$_x$
 are all renormalized and shifted, as compared to the DFT dispersions,
  leading to much smaller hole and electron pockets, as compared with calculations \cite{Watson2015a}.
  The renormalization values corresponding to the two main hole dispersions (with $d_{xz}$/$d_{xz}$ orbital character)
  are around 3-4 and hardly change for any compositions inside the nematic phase towards $x \sim$0.18,
  but they are reduced to a factor of $\sim $1-2.3 for FeS,
  suggesting that the suppression of electronic correlations occurs from FeSe towards FeS \cite{Reiss2017,Miao2017}.
  Additionally, the highly renormalized  $d_{xy}$ band, found at $\sim$50~meV below the Fermi level,
  remains relatively unaffected across the
nematic phase transition to $x \sim 0.18 $ and it cannot be
resolved for FeS  due to the disorder effects \cite{Reiss2017}.
The $d_{xy}$ hole band is notoriously difficult to observe in experiments
 due to matrix element effects and being  strongly incoherent in iron-chalcogenides \cite{Liu2015}
  but its dispersion can be revealed due to band mixing caused by the spin-orbit coupling effects \cite{Zhang2012b,Reiss2017}.
 As a function of $x$ substitution, the Fermi velocities increase by $\sim$10\% inside
 the nematic phase but more significantly outside towards FeS, reflecting the suppression
 of electronic correlations,  as shown in Fig.~\ref{Fig_ARPES_paramaters}e.

\begin{figure*}[htbp]
\centering
       \includegraphics[trim={0cm 0cm 0cm 0cm}, width=1\linewidth,clip=true]{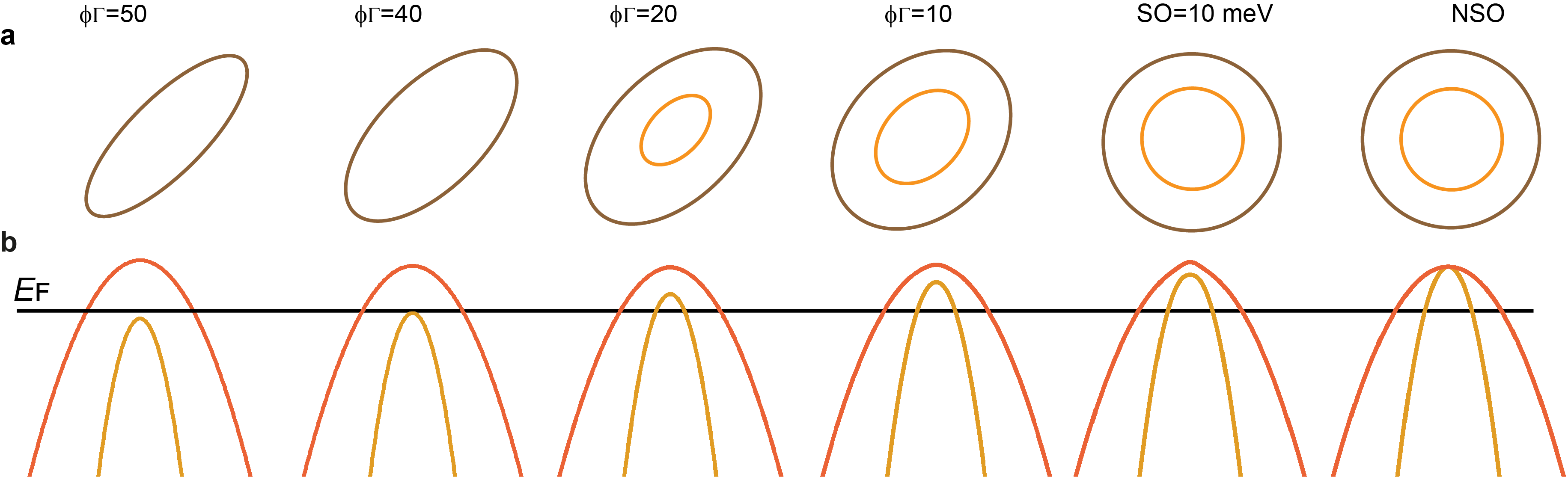}
              \includegraphics[trim={0cm 0cm 0cm 0cm}, width=1\linewidth,clip=true]{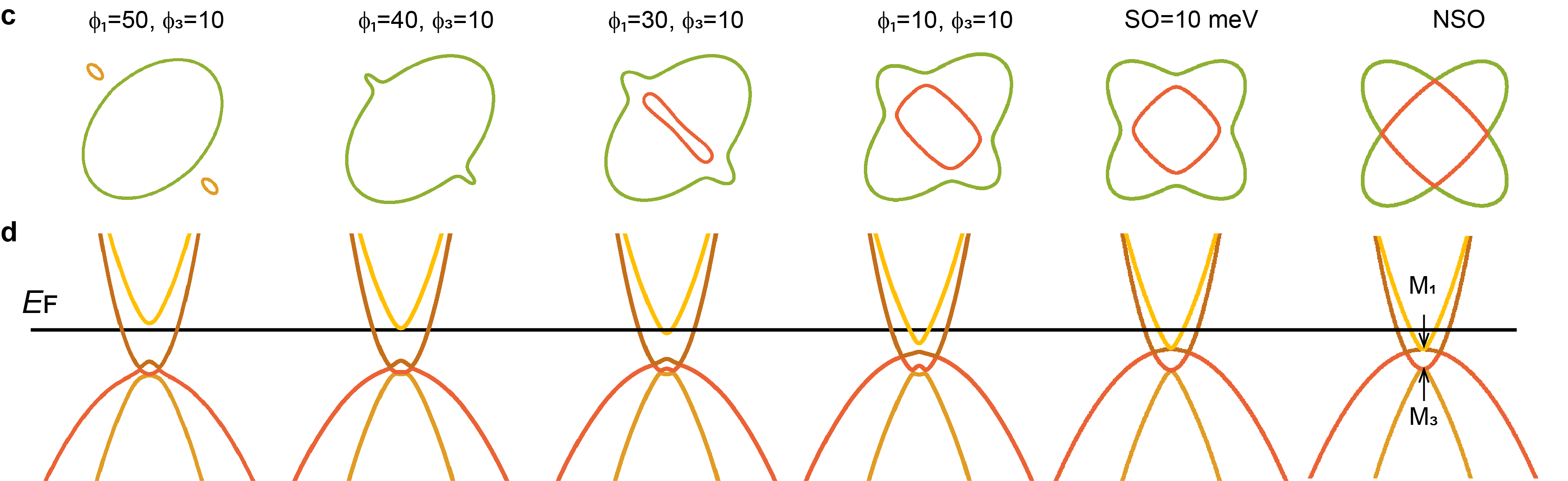}
       	 \caption{\textbf{Simulations of the effect of nematicity on the ARPES data.}
Simulations of the Fermi surface maps and band dispersions in a nematic and tetragonal phase, as described in detail in Ref.~\cite{Fernandes2014b,Christensen2020}.
This model parameters are adjusted to match the experimental data for the tetragonal $x$=0.18 in Fig.~\ref{Fig_ARPES}b and c
and the spin-orbit coupling is considered as being 10~meV.
The starting values used in simulations
are $\epsilon_1$ = -16, $\epsilon_3$=-35,
$m_1$ = 0.05,  $m_3$= $4\times 10^{-4}$,
$a_1$ = 964, $a_3$ = -2862, $v$ = -327,
 $p_1$ = -2589, $p_3$ = -589.
The simulations on the right side
assume the absence of spin-orbit coupling. The orbital order induced affects the electron and holes band dispersion in different way.
 a) At the $\Gamma$ point the band splitting is determined
both by the spin-orbit coupling
and nematicity, giving a band splitting of $\sqrt{\Delta^2_{SO}+\phi^2_{\Gamma}}$,
and the hole pockets become elongated in b). One of the inner hole pocket is pushed below the Fermi level.
c) At the M point the effect of nematicity is
influenced by the anisotropy of the on-site energies of the $d_{xz}$ ($\phi_1$)
and $d_{yz}$ orbitals,  anisotropic $d_{xy}$ hopping ($\phi_3$)
as well as the spin-orbit coupling \cite{Fernandes2014b} and the electron pockets changes shape significantly in d).
The in-plane maps illustrate the representation expected for a single domain sample.
In real experiments, the superposition of two different domains rotated by 90$^\circ$ could occur. }
 \label{Fig_simulations_ARPES}
\end{figure*}

The ARPES studies at the two high symmetry points using
different incident energies allows for the evaluation of the $k_{z}$ hole dispersion of the cylindrical Fermi
surfaces of FeSe$_{1-x}$S$_x$ and the sensitivity of the second inner hole band
to any band shifts inside the nematic phase.
The inner hole band, which forms a small 3D inner hole pocket around the Z point,
 is observed in the tetragonal phase of FeSe but it is pushed below the Fermi level
 inside the nematic phase \cite{Watson2015a}.
With sulphur substitution, the inner hole band is shifted gradually up and it crosses the Fermi level
only at the Z point from  $x \sim 0.11$ \cite{Watson2015c} and grows in size at $x \sim 0.18$.
The inner hole band does not cross the Fermi level at the $\Gamma$ point for any compositions
up to $x\sim 0.18$; however, it has been suggested that this pocket
could grow in size and become a two-dimensional cylinder
for FeS (Fig.~\ref{fig_phase_diagram}d and e) \cite{Reiss2017,Terashima2019,Miao2017}.
Interestingly, the Fermi surface of the tetragonal phase
for $x\sim 0.18$ is very similar to the Fermi surface
of FeSe at high temperature. Thus, there is
a direct correspondence between the temperature
and sulphur substitution in FeSe
 of the nematic electronic structure of FeSe$_{1-x}$S$_x$
 \cite{Watson2015c,Reiss2017}.

\subsection{\bf Simulations of the effect of nematicity}
 To understand the
 effect of nematicity and the orbital effects at each high symmetry point,
 simulations based on a model developed in Ref.~\cite{Fernandes2014b,Christensen2020}
 for a single domain sample are shown in  Fig.~\ref{Fig_simulations_ARPES}.
 The parameters for the simulations are adjusted to match the ARPES experimental data for
 $x=0.18$ in Fig.~\ref{Fig_ARPES}b and c \cite{Reiss2017}
 and the other variable are listed in Fig.~\ref{Fig_simulations_ARPES}.
In the tetragonal phase of FeSe$_{1-x}$S$_x$,
 the hole Fermi surface are expected to be circular and $C_4$ symmetric,
originating from the $d_{xz/yz}$ bands, as shown in Fig.~\ref{Fig_ARPES}e.
In the absence of nematicity the two hole dispersions at the centre of the Brillouin zone
are expected to be split only by the spin-orbit coupling \cite{Fernandes2014b}, as shown in Fig.~\ref{Fig_simulations_ARPES}a.
Experimentally, the band separation  gives a  spin-orbit of $\Delta_{SO}\sim$~13(3)~meV
for FeSe in the tetragonal phase at high temperatures
and for the tetragonal $x \sim 0.18$ at low temperatures
\cite{Watson2015a,Watson2017,Borisenko2015}.
As the nematic order is turned on, $\phi_{\Gamma}$, \cite{Fernandes2014b},
 the hole pocket is expected to become distorted and the
splitting between the two hole band dispersions
 increases, as shown in Figure~\ref{Fig_simulations_ARPES}.
In addition, the increase in the orbital ordering effects
moves the inner 3D hole pocket at the $Z$ point
completely below the Fermi level, as shown in Fig.~\ref{Fig_ARPES} \cite{Watson2015c,Suzuki2015}.
Experimentally, at low temperature FeSe has only one quasi-two dimensional hole Fermi surface
(compared with 3 predicted by the DFT calculations)
with an elliptical in-plane area at the high symmetry points, as shown in Fig.~\ref{Fig_ARPES}e.
The signature of this nematic electronic phase can be induced by orbital-ordering effects and
electronic interactions that can drive a Pomeranchuck instability of the Fermi surface \cite{Pomeranchuk1959}.
Other scenarios have been addressed theoretically in detail in other works \cite{Chubukov2016,Xing2017,Jiang2016,Scherer2017,Glasbrenner2015}.
Since samples of FeSe$_{1-x}$S$_x$ inside the nematic phase
can form twin domains rotated by 90$^{\circ}$ below $T_{\rm s}$
often ARPES experiments visualizes two superposed ellipses, as shown in Fig.~\ref{Fig_ARPES}e,
but only a single ellipse may be observed in detwinned measurements on FeSe \cite{Shimojima2014,Suzuki2015,Watson2017}.
As the orbital ordering is reduced with S substitution,
the splitting between the inner $d_{yz}$ bands and outer $d_{yz}$ hole band dispersion
is smaller and at low temperatures from $x \sim 0.11$,
the inner hole band crosses the Fermi level at the Z point, leading to
the formation of the small 3D hole pocket.
Inside the nematic phase, the in-plane  Fermi surfaces are highly anisotropic,
 indicated by the splitting of the two different $k_F$ values (obtained
 from the MDC cuts of two different domains), as shown in Figure~\ref{Fig_ARPES_paramaters}.
With increasing $x$,  in-plane Fermi surface becomes a circle
for both hole pockets for the tetragonal $x \sim 0.18$ but the cylindrical hole Fermi surface
has a strong $k_{z}$ dependence
 \cite{Reiss2017}.

\subsection{\bf Electron pockets of FeSe$_{1-x}$S$_x$}
Whereas the behaviour of the hole bands is well understood and consistent between different experimental reports,
the behaviour of the electron bands remains a highly debated subject.
The $P4/nmm$ unit cell of tetragonal FeSe includes two Fe sites which are related by a glide symmetry \cite{Fernandes2014b}
and ARPES measurements should detect  electron bands emerging from two-crossed ellipses  \cite{Watson2016,Fedorov2016},
similar to other systems, such as LiFeAs or NaFeAs \cite{Borisenko2015}.
The electron pockets suffer a significant change inside the nematic phase
and the relevance of different orbital contribution is still being debated.
At the corner of the Brillouin zone (M and A point) in the tetragonal phase,
there are two degenerate doublet states, M1 and M3, at the zone corner protected
by the space-group symmetry, even when spin-orbit coupling is taken into account \cite{Fernandes2014b},
as shown in Fig.~\ref{Fig_simulations_ARPES}c and d.
Therefore, any splitting and shifts of the bands
at M (or A)  would reduce the crystal symmetry in the presence of the spin-orbit coupling.
The nematic order can be triggered by the development of the anisotropy
in the on-site energies of the $d_{xz}$ and $d_{yz}$ orbitals ($\phi_1$ term)
and anisotropic $d_{xy}$ hopping ($\phi_3$ term) \cite{Fernandes2014b}.
The two-crossed ellipse, corresponding to the electron pockets (Fig.~\ref{Fig_ARPES}h), are expected to have a finite splitting between the inner and
outer orbits, due to the spin-orbit coupling in the tetragonal phase, as shown in Fig.~\ref{Fig_simulations_ARPES}c.
Inside the nematic phase, by increasing both the $\phi_1$ and $\phi_3$, the degeneracy at the M1 and M3
points are lifted and the bands split apart; this promotes the in-plane distortion of the Fermi surface along
its longest axis (Fig.~\ref{Fig_simulations_ARPES}c,d). Further increasing $\phi_1$, which is related to orbital order induced
by the increase orbital polarization of the $d_{xz}$ versus $d_{yz}$ bands, the inner electron band is pushed up
and eventually it can disappear; thus only a single electron pocket is present, as shown in Fig.~\ref{Fig_simulations_ARPES}d.
Furthermore, orbitally-induced shifts could shrink the electron pocket along one direction, transforming it
into two small Fermi pockets, whereas along the other direction, the electron pocket is enlarged into a peanut shape \cite{Zhang2016,Watson2017}.
Indeed, at low temperatures, the band giving rise
to the inner electron pocket at the M point
 is very close to the Fermi level about 3~meV for FeSe (within experimental resolution of 3~meV).
 This proximity creates the conditions for a topological transition of the electron pocket into a peanut or Dirac-like crossing,
 under other perturbations, such as applied strain \cite{Watson2017strain,Cai2020},
 as found for thin films of FeSe under internal strain from
the substrate \cite{Zhang2016}.

Experimentally, in the tetragonal phase the inner electron band dispersions at M (or A) are expected
to have $d_{xz}/d_{yz}$ orbital character when probed along the diagonal of the Brillouin zone, as shown in Figure~\ref{Fig_ARPES}h.
The outer electron band with $d_{xy}$ orbital character is harder to observe
due to matrix element effects and the incident energy used (Fig.~\ref{Fig_ARPES}c and j).
This behaviour is detected for the tetragonal  FeSe$_{1-x}$S$_x$  with $x=0.18 $ shown in
Fig.~\ref{Fig_ARPES}(c) \cite{Reiss2017}
and FeSe above $T_{\rm s}$  \cite{Watson2015a,Watson2015c,Watson2016}.
In the nematic phase, the changes for the electron bands
are drastic, with M1 point shifting up whereas the M3 shifts down, and additional splitting
could take place around these two degenerate points, as shown in Fig.~\ref{Fig_ARPES}c.
The energy separation between the two intense features at the M point below $T_s$ (EDC cuts),
is defined  by $\Delta_M$ which is $\sim 50$~meV for bulk FeSe
\cite{Shimojima2014,Watson2015a,Zhang2015}.
This splitting is much larger than what would be expected from DFT calculations
 simply taking into account its small orthorhombic distortion ($\sim 5$~meV) \cite{Watson2015a}.
 The elongated directions of the elliptical Fermi surfaces at the M point are rotated by 90$^\circ$ degrees with respect to that
 at the $\Gamma$ point due to the momentum-dependent sign-changing orbital polarization,
where the $d_{xz}$ band shifts upward at the $\Gamma$ point but downward at the M point \cite{Suzuki2015,Zhang2016b}.
  Interestingly, 20~meV already separates the
  two doublets in the tetragonal phase at the M point,  in the absence of nematicity,
  for the tetragonal system with $x=0.18 $ (see Fig.~\ref{Fig_ARPES}g),
and FeSe  at high temperatures \cite{Watson2017}.
This implies that the energy
scale of the nematic order could be smaller that 50~meV,
as shown in Fig.~\ref{Fig_ARPES_paramaters}a \cite{Coldea2017}.

A direct signature of the nematicity is the in-plane distortion of the Fermi surface.
Inside the nematic phase for the electron pockets
this can be related to the development of the orbital polarisation $\Delta{}n=n_{xz}-n_{yz}$.
The orbitally dependent band shifts cause the inner sections of the electron pockets with $d_{yz}$ orbital character
to contract whereas the $d_{xz}$ sections to expand,  but forming a cross-shape due to effect of sample twinning,
 as shown in Fig.~\ref{Fig_ARPES}f.
 The degree of anisotropy of the Fermi surface
 can be related to $(k_F-k_{F0}$)/$k_{F0}$ \cite{Watson2015c}.
where the $k_F$-vector is that corresponding to the inner $d_{yz}$ portion of the electron pocket,
 and $k_{F0}$ is the Fermi $k$-vector in the tetragonal phase for each compound.
Figure~\ref{Fig_ARPES_paramaters}b shows the evolution of the Fermi surface elongation with $x$ substitution
and indicates that the nematic phase is responsible for this in-plane distortions,
which is completely suppressed in the tetragonal phase.

The presence of both highly elongated and isotropic  Fermi surfaces of FeSe$_{1-x}$S$_x$
is likely to significantly influence other measurements.
STM studies shows highly anisotropic QPI patterns inside the nematic state, becoming
isotropic in the tetragonal phase \cite{Sprau2016,Hanaguri2018}.
The resulting QPI spectra
 exhibit electron-like and hole-like dispersions along different directions
 ($q_a$ and $q_b$, respectively)  corresponding to the intraband back-scatterings in the electron bands at the
 Brillouin zone corner and in the hole band at the zone center, respectively.
 Thus,  the QPI spectra reflect  the evolution of the scattering processes
  across the series FeSe$_{1-x}$S$_x$.

\subsection{Comparison between ARPES and QPI}
To clarify the qualitative relation between the QPI branches and the band structure,
the scattering $q$ vectors from the intraband backscattering can be compared
with the Fermi wavevector extracted directly from ARPES dispersions at the Fermi level, as shown in Fig.~\ref{Fig_ARPES_paramaters}c.
The Fermi momenta of FeSe of a distorted deformed Fermi cylinder can be estimated from
the scattering vectors at zero energy ($q/2 \sim k_{\rm F}$)  to be
$\sim 0.05$ and 0.08 \AA$^{-1}$ for the hole band
and $\sim 0.04 $~\AA$^{-1}$ for the electron band \cite{Hanaguri2018,Shibauchi2020}.
On the other hand, the Fermi wavevector from ARPES for FeSe
for the elliptical hole pocket  at the $\Gamma$ point
varies between $\sim 0.035$ to 0.08 \AA$^{-1}$, \cite{Watson2015a,Watson2015c}
but is larger at the Z point ($\sim 0.1$ and 0.15 \AA$^{-1}$),
 as shown in Fig.~\ref{Fig_ARPES_paramaters}c.
These values are close to those from laser ARPES data (which are usually measured around 7~eV  which corresponds to a $k_z$ position
closer to the $\Gamma$ point),  varying between
$0.036~\pi /a  \sim 0.038$ \AA$^{-1}$ and $0.11~\pi /a \sim 0.092 $ \AA$^{-1}$.
Thus, the  resulting elongated hole ellipse of FeSe, with a high aspect ratio ($\sim$3), is one of the most anisotropic Fermi surfaces
 among all the iron-based superconductors \cite{Liu2018}.
Based on the direct comparison between ARPES and QPI data on FeSe$_{1-x}$S$_x$,
the scattering vectors in QPI directly correspond to the in-plane Fermi vectors at the centre of the Brillouin zone ($k_z=0$).
They are less sensitive to $k_z$ dependent scattering processes outside of this plane,
despite recent theoretical suggestions for FeSe \cite{Rhodes2019},
as the $k_{\rm F}$ vectors at $Z$ are much larger than the scattering vectors  ($q/2 $) extracted from QPI,
as shown in Fig.~\ref{Fig_ARPES_paramaters}c.
In the case of the electrons pockets, the QPI scattering vector is close to those of the small inner electron wave vector
($\sim 0.02(1)$ \AA$^{-1}$) rather than to the long elongated axis of the ellipse (0.14(1) \AA$^{-1}$),
found in ARPES \cite{Watson2016,Coldea2017}.
Furthermore, the estimated Fermi energies from the QPI dispersions,
for the hole bands are of 10-20~meV, \cite{Hanaguri2018}
in good agreement with the top of the hole band at the $\Gamma$ point
of $\sim 17$~meV from ARPES \cite{Watson2015a,Coldea2017}.
Laser ARPES data  (measured away from a high-symmetry point)
give slightly lower values of $\sim $6.7 meV or 10~meV for the hole band \cite{Liu2018,Hashimoto2018}.
For the  the electron bands
the Fermi energies of 5-10~meV from QPI
are close to the ~3-5~meV corresponding to the inner electron band
dispersion \cite{Watson2017,Coldea2017}.
With increasing $x$,  the bottom of the inner electron bands at the M point is pushed
lower below the Fermi level from about 3~meV
 towards 15~meV for $x$=0.18 \cite{Reiss2017,MacFarquharson2020}.
 The outer hole band crossing at the Z point
lies around 25(3)~meV from $x$=0 to $x$=0.18,
but  decrease more significantly for the hole point
at the $\Gamma$ point (Fig.~\ref{Fig_ARPES_paramaters}d).
These shifts bring the cylindrical hole band into the regime to undergo a possible Lifshitz tranistion
at the nematic end point, as suggested by quantum oscillations \cite{Coldea2019}.
The QPI dispersion associated with the long hole axis scattering vector ($\Gamma$ point)
  was suggested to disappear close to $x \sim $0.11,
   whereas the short axis scattering vector was suggested
   to  increase towards $x \sim 0.25$, as shown in Fig.~\ref{Fig_ARPES_paramaters}c,
 after Ref.~\cite{Hanaguri2018}.
However, there is no evidence for the disappearance of hole bands up to $x$=0.18
in ARPES data (Fig.~\ref{Fig_ARPES_paramaters}c).
 One can envisage that the two scattering vectors, ($q_1$ and $q_2$)
 would merge into one value for a isotropic Fermi surface close to $x \sim 0.17$.
Thus, as the system becomes isotropic a single hole dispersion is visible in the QPI
and the scattering vectors are similar to those from ARPES at the $\Gamma$ point.
As QPI is less sensitive to $k_z$ dependent scattering processes,
 the small 3D hole band at the $Z$ point in Fig.~\ref{Fig_ARPES_paramaters}c is not detected \cite{Hanaguri2018}.
 Further theoretical work will be needed to reconcile quantitative
 features obtained from ARPES and QPI for FeSe$_{1-x}$S$_x$.

\begin{figure*}[htbp]
	\centering
   \includegraphics[trim={0cm 0cm 0cm 12cm}, width=1\linewidth,clip=true]{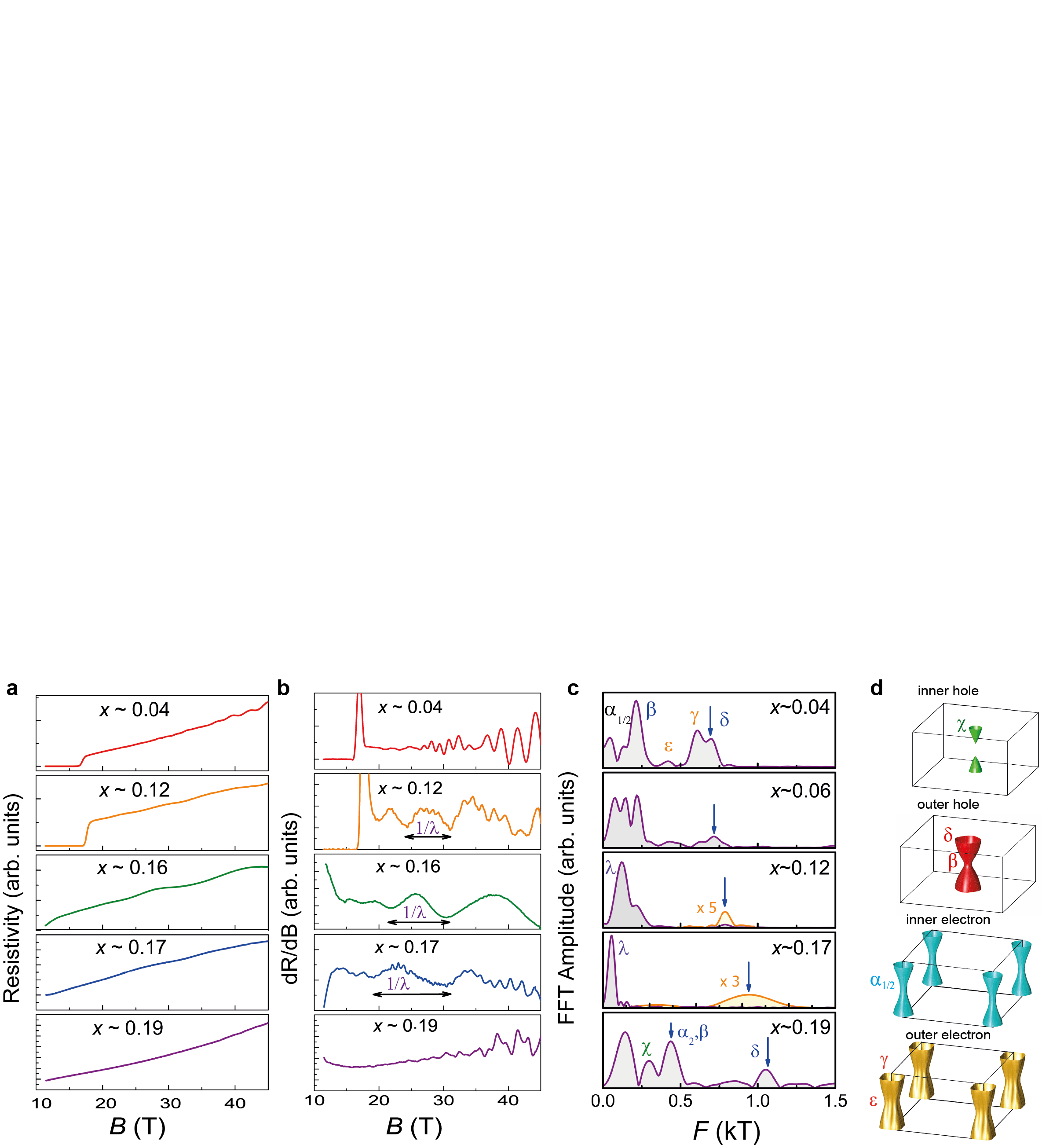}
	\caption{\textbf{Quantum oscillations in FeSe$_{1-x}$S$_x$.}
a) The in-plane resistance $R_{xx}(B)$ as a function of magnetic field $B$ for different compositions, $x$,
measured at $T \sim 0.35$~K.
b) The oscillatory part of the resistivity visualized better by the first derivative of resistance as a function of magnetic field from a-f) at the lowest measured temperature. The horizontal line indicates  the period of the low-frequency oscillations ( $\sim 1/\lambda$).
c) The frequency spectra of the oscillatory signal ($\Delta \rho_{osc}$/$\rho$) obtained
by subtracting a polynomial background and using a fast Fourier transform (FFT).
A multiplied FFT spectrum is used to emphasize the weak high frequency $\delta$ frequency for
certain $x$ compositions  indicated by vertical arrows.
The proposed Fermi surface and the different extremal areas for FeSe$_{1-x}$S$_x$ obtained by shrinking the
calculated tetragonal Fermi surface of FeSe from Fig.~\ref{fig_phase_diagram}e.
Frequencies below 200~T cannot be reliably assigned due to a possible overlap
of at least 3 different small frequencies ($\alpha_1$, $\alpha_2$ and $\chi$).
These graphs are adapted from Ref.~\cite{Coldea2019}.}
	\label{fig_quantum_oscillations}
\end{figure*}
\section{\bf Quantum oscillations in high magnetic fields }
A powerful technique to access directly the Fermi surface of  FeSe$_{1-x}$S$_x$
is via quantum oscillations in very high magnetic fields and at low temperatures below 1.5~K
\cite{Terashima2014,Watson2015a,Coldea2019,TerashimaFeS}.
Quantum oscillations originate from the  oscillations in the density of states
in the presence of the Landau quantisation of a metallic system in an applied magnetic field.
The quantum oscillations are periodic in $1/B$ and the
frequency of these oscillations relates directly to extremal areas of the Fermi surface via Osanger relation
($F_i=\hbar/2\pi e \cdot A_{k_z,i}$ with frequencies
in Tesla $\sim 10^{-16} $ (\AA$^{-2}$) of the cross-section area of each orbit).
For a slightly-warped cylindrical Fermi surface
two frequencies would be observed at the centre ($k_z=0$)
and the top of the Brillouin zone ($k_z=\pi/c$), and in the case of
twinned crystals the cross section areas of different domains would coincide.
Quantum oscillations are normally observed only in clean single crystals
as the cyclotron energy which separates Landau levels needs to be larger than
the broadening of the levels $\hbar/\tau$ due to scattering.
Quantum oscillations have been observed for all $x$ compositions of FeSe$_{1-x}$S$_x$
\cite{Coldea2019,Terashima2014,Terashima2019,Audouard2014}.
The isoelectronic substitution result
in relatively similar mean free paths
(using the Dingle analysis for the maximum hole band orbit \cite{Shoenberg1984}),
having values of $\ell \sim 277$(35)~\AA~ for FeSe
and $\ell \sim  283$(20)~\AA~ for $x \sim 0.19$
 \cite{Coldea2019}.
Besides impurity scattering effects, the amplitude of the quantum oscillations is significantly suppressed
for heavier quasiparticle masses as a result of the smearing of the Landau levels by the Fermi-Dirac distribution
and often heavier masses cannot be observed.

\subsection{\bf Comparison between ARPES and quantum oscillations}
As compared with ARPES, quantum oscillations are insensitive
to surface states and the signal is  dominated by the bulk response thus giving an unambiguous
probe of the bulk Fermi surface. Furthermore, they have a much better $k$-space
resolution of 10$^{−3}$ of the area of the Brillouin zone
that allows very accurate determination of the cross-section orbits on the Fermi surface, for a particular magnetic field orientation.
However, the location of the orbits in the $k$-space is not easily known for multiband systems
(Fig.~\ref{fig_phase_diagram}d-g)
making the assignment of the potential frequencies for a multi-band system difficult.
In these circumstances,  the angular dependence
of the observed orbits is used as a guide to assign the different orbits to Fermi surfaces
as the minimum and maximum orbits will have different angular dependencies
and it is expected that the cyclotron effective mass for the same band is likely to have similar values
\cite{Watson2015a,Terashima2014}.
Even in twinned samples, the quantum oscillations are likely to be unaffected
as the cross-section areas originating from different domains would be the same,
however, any differences only be noticeable at very high rotation
angles where the quantum oscillation amplitude disappears.
The experimental Fermi surface of FeSe$_{1-x}$S$_x$ could potentially have
four different sheets, that could generate up to seven or eight extremal orbits at the high symmetry points due the strong $k_z$ dependence,
as shown in Fig.~\ref{fig_quantum_oscillations}d.
In a system with many similar small orbits (with $k_F$ values below 0.08~\AA)
the expected frequencies would be found below  200~T.
A clear separation between individual small frequencies is hampered
by  the  limited magnetic field window (20-45~T) caused
by the presence of superconductivity and large upper critical field ($< 20$~T) (Fig.~\ref{fig_quantum_oscillations}a.
This low frequency region is also affected by extrinsic effects in a fast Fourier transform (Fig.~\ref{fig_quantum_oscillations}c,
such as the $1/f$ noise and the peak created by a background polynomial,
making any reliable assignment of the small frequencies difficult.

An interesting insight
into the origin of the quantum oscillation amplitudes
was provided by magnetotransport and Hall effect in ultra-high magnetic fields up to 90~T in FeSe  \cite{Watson2015b}.
By comparing the changes in the relative amplitudes of the quantum oscillations of the
$\rho_{xx} $ and $\rho_{xy} $ components, and considering the positive sign of the high-field Hall signal at very
low temperatures  \cite{Watson2015b}, the mobile carriers were assigned to the hole band
 ($\beta$ and $\delta$ orbits) ~\cite{Terashima2014,Audouard2014,Watson2015a}.
The frequencies of quantum oscillations assigned to the quasi-two dimensional hole cylinder of FeSe
are 220~T for $\beta$ orbit at the $\Gamma$ point
and 660~T for the $\delta$ orbit at the Z point, when magnetic field $B||c$.
The cyclotron effective masses associated with these two orbits are around $4.5(5)~m_{\rm e}$,
in good agreement between different studies \cite{Terashima2014,Watson2015a,Coldea2017}.
The estimated frequencies of the hole pockets
using the $k_F$ values from ARPES data (Fig.~\ref{Fig_ARPES_paramaters}c)
are consistently smaller than those assigned in quantum oscillations
(by $\sim 150$~T or 24\% smaller for the $\delta$ orbit at the Z point and $\sim 100$~T or 50\% smaller
  for the $\beta$ orbit at the $\Gamma$ point for FeSe).
One obvious difference between the two techniques
 is  related to sensitivity to surface states in ARPES, compared with bulk,
 that is normally probed by quantum oscillations.
 ARPES resolution, $k_{z}$ dependence and the energy and momentum integrations is expected
to affect the precise $k_{\rm F}$ values.
Variation of the values for the top of the hole band at the $\Gamma$ point of 6.7-15 ~meV
is found between different reports for FeSe \cite{Hashimoto2018,Watson2015a},
and this will affect the precise determination of the $k_{\rm F}$ values.
On the other hand, the high frequency values from quantum oscillations
have a much better agreement between different reports \cite{Terashima2014,Watson2015a,Audouard2014}.
Quantum oscillations are measured below 1~K in high magnetic
field above 20~T whereas ARPES is measured in zero field above 10~K ($\sim $~1meV).
As inner pockets of  FeSe$_{1-x}$S$_x$ are small, magnetic field
could induce additional spin-polarization of the Fermi surfaces
in very high magnetic fields (3-4~meV).
Another discrepancy between quantum oscillations and ARPES is
related to the orbitally averaged effective masses that are
larger for the outer electron bands ($\sim 7 m_{\rm e}$) compared with the hole bands ($\sim 4.5 m_{\rm e}$)
in quantum oscillations \cite{Watson2015a,Terashima2014}.
In contrast, the Fermi velocities extracted
from ARPES are  larger at the A point (0.66 eV \AA) compared with
hole bands at the Z point (0.4-0.5 eV \AA) \cite{Watson2017strain}.
However, the velocities in ARPES are extracted
for only one of the highly symmetry direction
and the values are not orbitally averaged (Fig.~\ref{Fig_ARPES}).

Quantum oscillations in iron-based superconductors
detect clearly electron Fermi surfaces with lighter effective masses
in LaFePO \cite{Coldea2008}, LiFe(As/P) \cite{Putzke2012} and BaFe$_2$As$_2$ \cite{Shishido2010}.
These orbits originate from inner and outer quasi-two dimensional cylinders
due to the finite spin-orbit coupling, as depicted  for FeSe$_{1-x}$S$_x$ in Fig.~\ref{Fig_simulations_ARPES}c.
In FeSe, the orbital differentiation is  much more pronounced than in iron pnictides, as the $d_{xy}$ band is
involved in the formation of outer flower-shaped electron orbit (Fig.~\ref{Fig_ARPES}h).
This would lead to a much heavier orbitally averaged cyclotron mass of $\sim 7$~m$_e$,
 associated with the outer electron orbit around the A point
 ($\gamma$ orbit of $\sim 560$~T)  for  FeSe, as shown in Fig.~\ref{fig_quantum_oscillations_meff}a \cite{Terashima2014,Watson2015a}.
Based on $k_F$ values at the A point determined from ARPES with Fermi vector values of 0.03(1) and 0.19(1) \AA$^{-1}$ \cite{Watson2017strain},
the area of a flower-shaped orbit would be $\sim 350$~T (or $\sim 35$\% of $\gamma$ orbit) whereas for a single ellipse pocket
reaches only $\sim 190$~T, which is or $\sim 66$\% smaller than the  $\gamma$ orbit from quantum oscillations.
 In ultra high magnetic fields, potential breakdown orbits could be generated by
 tunneling across the gaps created by the spin-orbit coupling,
 but the necessary magnetic fields are likely to be very large
 and the orbits would be smaller than that of the flower-shaped orbit \cite{Putzke2011comment}.
The nematicity has a drastic effect on the electron bands
and it can lead to highly elongated pockets with  a very small inner electron band, as shown in Figure~\ref{Fig_simulations_ARPES}c.
At 13~K, the inner band at the M point gets very close to the Fermi level within 3~meV for FeSe (within the experimental resolution).
Thus, any small changes in the band positions relative to the chemical potential (1-2~meV) that
could occur at low temperatures below 1.5~K and in high magnetic fields could potentially
push the inner electron bands above the Fermi level at the M point
and lead to single elliptical orbit or an elliptical pocket and two tiny electron pockets,
as shown in Figure~\ref{Fig_simulations_ARPES}c.

Recent studies promotes the idea that FeSe would only have a single electron pocket in the corner
of the Brillouin zone. For a {\it peanut}-like pocket at the A point \cite{Watson2017strain}
its area is almost a factor 3 smaller than the $\gamma$ pocket in quantum oscillations,
the change compensation of the system would be lost and the magnetotransport data of FeSe cannot be explained \cite{Watson2015b}.
The proximity of the inner electron band to the Fermi level is highly sensitive
to small energetic alterations within experimental resolution ($\sim 3$~meV),
 any differences in Fe stoichiometry, surface effects
 or the possible changes that can occur under applied uniaxial stress,
 as found for thin films of FeSe under strain from a substrate \cite{Zhang2016b}).
 Thus, different scenarios related to the fate and the number of the electron pockets  (Fig.~\ref{Fig_simulations_ARPES}c)
need to be considered, besides other theoretical reasons
 \cite{Watson2017strain,Rhodes2020,Yi2019,Huh2020,Rhodes2020nonlocal}.

Figure~\ref{fig_quantum_oscillations}c show the complex fast Fourier spectra of FeSe$_{1-x}$S$_x$
 due to the presence of multiple small Fermi surfaces areas.
The signature associated with the inner electron band in quantum oscillations
would be a peak in the Fast Fourier transform below 100~T.
Previously, it was assumed that FeSe has a single cylindrical electron pocket, with areas varying between
 50~T pocket for its minimum and  550~T ($\gamma$) for its maximum  \cite{Terashima2014}.
 This variation would suggest a much more warped Fermi surface cylinder
for the electron band (factor 10 between the two high symmetry areas),
as compared with the hole band, for FeSe, in disagreement
with the $k_z$ dependence determined from ARPES studies \cite{Watson2015a,Coldea2017,Watson2017strain}.
The quantum oscillations spectra could assign the lowest frequency below 100~T
to the inner quasi-two dimensional electron pocket,
whereas $\gamma$  and $\epsilon $ around 440~T could correspond to the outer electron band
 (Figure~\ref{fig_quantum_oscillations}d) \cite{Watson2015a,Coldea2017}.
 A small inner electron pocket is not easy to observe using spectroscopic techniques, nor does it have
a large contribution to the density of states, but it plays an important role in magnetotransport
due to its high mobility  \cite{Watson2015c}.
With sulphur substitution for small $x<0.09$, quantum oscillations
show similar features to those of FeSe,  as shown in Figure~\ref{fig_quantum_oscillations}c.
As the nematic effects are progressively removed, the inner electron orbits would increase in size,
reaching a value of 200~T in FeS \cite{Terashima2019}.

\subsection{\bf Evolution of Fermi surface areas of FeSe$_{1-x}$S$_x$ }
\begin{figure*}[htbp]
	\centering
   \includegraphics[trim={0cm 0cm 0cm 0cm}, width=1\linewidth,clip=true]{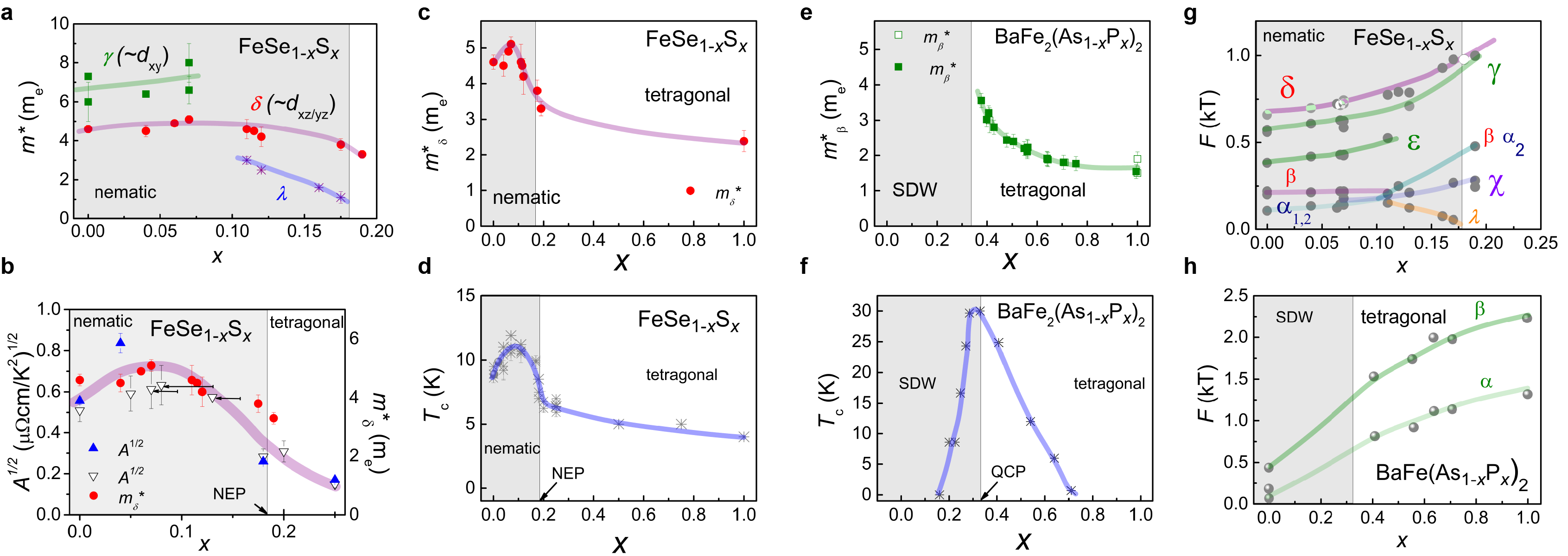}
  	\caption{\textbf{Electronic correlations of FeSe$_{1-x}$S$_x$ compared with BaFe$_2$(As$_x$P$_{1-x}$)$_2$.}
  a)  The quasiparticle effective masses of the high frequencies, $\gamma$ and $\delta$ and the dominant low frequency, $\lambda$.
  Solid lines are guides to the eyes.
 Grey areas indicate either the nematic phase for FeSe$_{1-x}$S$_x$  or the spin-density wave (SDW) for
 BaFe$_2$(As$_x$P$_{1-x}$)$_2$.
  b) Comparison between the Fermi liquid coefficient $A$ and the effective masses for different compositions of FeSe$_{1-x}$S$_x$.
		The band masses of the $\delta$ orbit (from Ref.~\cite{Coldea2019})
are compared to $A^{1/2}$, extracted from the low temperatures Fermi-liquid behaviour,
and shown as solid triangles (after Ref.~\cite{Bristow2020}).
Data shown as open triangles were reported in Ref.\cite{Licciardello2019} using nominal $x_{nom}$ values and
they are shifted in $x$ to smaller values (as indicated by horizontal arrows) to match
the real concentration based on the values of $T_s$, as reported previously \cite{Coldea2019,Hosoi2016}.
The nematic end point (NEP) occurs close to $x \sim 0.175(5)$  \cite{Coldea2019,Hosoi2016} and
the solid thick lines are guides to the eye.
The comparison between the effective mass of maximum orbit of the outer hole (at the Z point), $\delta$, in
FeSe$_{1-x}$S$_x$ in c) and the maximum orbit $\beta$ of the outer electron band (at the A point)
in  tetragonal phase of BaFe$_2$(As$_x$P$_{1-x}$)$_2$ \cite{Shishido2010,Walmsley2013}  in e)
together with the evolution of the superconducting critical temperature, $T_c$ in d) and f), respectively.
The quantum oscillations data for FeS are from Ref.~\cite{TerashimaFeS}.
A qualitative correlation is found between the $T_c$ and the electronic correlations
for the two systems, but superconductivity in the vicinity of the SDW is a fact 3 stronger than in FeSe$_{1-x}$S$_x$.
Superconductivity is enhanced at the magnetic critical point in BaFe$_2$(As$_x$P$_{1-x}$)$_2$ but this does not occur
at the nematic end point in FeSe$_{1-x}$S$_x$, suggesting that the nematic fluctuations are not the main driver for superconductivity.
Instead, a local peak in $T_c$ inside the nematic phase may signify an enhanced low-energy spin fluctuation in this regime \cite{Wiecki2018}.
The evolution of the multiple observed quantum oscillations frequencies of
  FeSe$_{1-x}$S$_x$ in   g)  compared with those of BaFe$_2$(As$_x$P$_{1-x}$)$_2$ in h).
The lines are guides to the eye to indicate the potential origin of the observed frequencies.
Frequencies of FeS are also larger \cite{TerashimaFeS}.
Pronounced Fermi surface shrinking is observed in both systems as the electronic correlations are enhanced.
   }
	\label{fig_quantum_oscillations_meff}
\end{figure*}

The overall evolution of the Fermi surface of FeSe$_{1-x}$S$_x$ implies that the majority of the cross-sectional areas
expand as a function of chemical pressure, in particular the maximum orbits located at the top of the Brillouin zone,
as shown in Fig.~\ref{fig_quantum_oscillations_meff}g \cite{Coldea2019}.
For the outer hole band ($\delta$ orbit), the increase in areas reflects the transition from an in-plane anisotropic
to isotropic Fermi surface, as the ellipse transforms into a circle, and as  the in-plane areas increase
due to changes in the lattice parameters  \cite{Watson2015c,Reiss2017,Coldea2017}.
These trends are in contrast to the small Fermi surfaces observed under applied pressure in FeSe,
 suggested to result from Fermi surface reconstruction inside the spin-density phase  \cite{Terashima2015}.
However, the Fermi surfaces of FeSe$_{1-x}$S$_x$ are severely reduced in size compared
  with those predicted by DFT calculations (varying from a factor of 5 for FeSe towards  a factor 3 for $x \sim 0.17$).
   This shrinking  is an important consequence of strong orbitally-dependent inter- and intra-band electronic interactions, significantly large in iron chalcogenides \cite{Fanfarillo2016,Watson2015a}, but also found in many iron-based superconductors \cite{Shishido2010,Coldea2008}.
  These effects are suppressed once the bandwidth increases with sulphur substitution towards FeS \cite{Reiss2017}
  or with phosphorus substitution in BaFe$_2$(As$_x$P$_{1-x}$)$_2$,
  as shown in Fig.~\ref{fig_quantum_oscillations_meff}g and h \cite{Shishido2010}.
The largest orbit detected in FeS is almost a factor 2 larger than for $x\sim 0.19.$ \cite{Terashima2019} but
it is still a factor 2 smaller than that predicted by band structure, and band-shifts of 0.1~eV are required
to bring experiment in agreement with DFT calculations \cite{Terashima2019}.
The Fermi energies estimated from quantum oscillations of FeS
have significantly increased to $27-102$~meV \cite{Terashima2019},
compared with 3-18~meV estimated for FeSe
\cite{Terashima2014}.

Transport measurements in a multi-band system like FeSe$_{1-x}$S$_x$  are normally dominated by
the pockets with the highest mobility carriers in a parallel resistor model.
The magnetoresistance at low temperatures
shows a prominent low frequency oscillation from $x=0.12$ towards NEP (Fig.~\ref{fig_quantum_oscillations}a and b) \cite{Coldea2019}.
Outside the nematic phase,  the background magnetoresistance
is almost quadratic in magnetic field and the dominant low-frequency oscillation has disappeared (Fig.~\ref{fig_quantum_oscillations}).
This dominant low frequency is not detected at higher sulphur substitution or higher pressures
beyond the nematic end point inside the tetragonal phase \cite{Coldea2019,Reiss2020}.
In quantum oscillations, the disappearance of a frequency
could be linked to a possible Liftshitz transition, which
is a topological change of the Fermi surface
in which the neck of a quasi-two dimensional is disconnected while the top of
the cylinder expands such that the volume remains the same \cite{Lifshitz1956}.
ARPES data indicate that a small inner 3D hole pocket centered at Z is
expected to emerge from $x \sim 0.11$, as shown in Fig.~\ref{Fig_ARPES}.
This small 3D pocket is supposed to grow in size with $x$, rather than to disappear.
However, in high magnetic fields, this 3D hole pocket could become heavily spin-polarized
and, therefore, one of its polarized sheet could disappear at the nematic phase boundaries.
Another scenario can rely on the strong increase in the interlayer  warping as a function of chemical or applied pressure,
as the conducting layers come closer together when $c$ axis decreases.
 DFT calculations of FeSe and FeS
show indeed that the hole bands are highly sensitive to the position of the chalcogen atom above the Fe plane
\cite{Watson2015a,Reiss2017,Watson2016}.
ARPES studies for the tetragonal $x\sim {0.18}$ suggest that
 the hole band at the $\Gamma$ point  is smaller compared with FeSe
 and the top of the band is about $~5$~meV above the Fermi level,
 as shown in  Fig.~\ref{Fig_ARPES_paramaters}d \cite{Coldea2017,Reiss2017}.
 Thus, the orbit associated with the hole band at the $\Gamma$
 point could be a prime candidate for the observed
disappearance of a significant frequency in quantum oscillations.
 Other scenarios could invoke magnetic field-induced Lifshitz transitions
 affecting the bands with very small Fermi energies,
 such as the inner hole and electron bands,
 that are comparable to the Zeeman energy (3-4~meV)  \cite{Ptok2019}.
 Multi-band interference effects
as well as oscillations of the chemical potential
could be considered as other potential theoretical avenues to understand
these effects in magnetic fields \cite{Kartsovnik2004}.

\subsection{\bf Electronic correlations at a putative nematic critical point}
The cyclotron-averaged effective masses of the quasiparticles
for each extremal orbit can be extracted from the temperature dependence of the
amplitude of the quantum oscillations \cite{Shoenberg,Coldea2019}.
 The quasiparticle masses associated with the largest hole orbit $\delta$
 around the Z point increase slowly from 4.3(3)$m_{\rm e}$
 towards a local maximum around $x <0.11$ before
 the values continue to decrease outside the nematic phase
 to around $3.2(5) m_e$ for $x \sim 0.19$.
 In the end compound FeS, quantum oscillations have revealed very light effective
 masses ranging from $0.6-2.1m_e$ \cite{Terashima2019,TerashimaFeS}.
 The overall trends shows that cyclotron masses
are larger inside the nematic phase of FeSe$_{1-x}$S$_x$
but they are getting lighter with the increasing bandwidth \cite{Miao2017}.
The reduction of the electronic correlations towards
FeS is supported by the enhanced velocities from ARPES,
shown in Fig.~\ref{Fig_ARPES_paramaters}e \cite{Reiss2017}.
The effective mass of the prominent small frequency oscillation ($\lambda$) is small below $2m_e$.
 Due to its heavy mass and its possible proximity to the $\delta$ orbit,
 the $\gamma$ orbit (with some orbitally averaged $d_{xy}$ character)
  cannot be detected over the entire range but it is expected to
 follow similar trends to the hole bands effective mass ($\delta$) and
 to the electronic contribution to the specific heat (Fig.~\ref{fig_quantum_oscillations_meff}a) \cite{Coldea2019,Sato2018}.

 The nematic state of FeSe$_{1-x}$S$_x$
 is a correlated electronic state based on the quasiparticle effective masses.
  Interestingly, the electronic correlations  assigned to the orbits with predominant $d_{xz}/d_{yz}$  character (outer hole band, $\delta$)
 follow similar trends as $T_c$, as shown in Figure~\ref{fig_quantum_oscillations_meff}c and d,
suggesting that this quasi-two dimensional hole band is likely to play a dominant role in the pairing mechanism.
  The trends in the effective masses are in good agreement with those from
  specific heat studies on FeSe$_{1-x}$S$_x$ that show a slight increase in the Sommerfeld coefficient
  (7-9 mJ/mol K) inside the nematic phase
before being smoothly suppressed, without  any enhancement at the nematic end point
  \cite{Wang2016Meingast,Sato2018,Abdel-Hafiez2015}.
  Additionally, the Fermi liquid behaviour $A^{1/2}$ coefficient extracted from the low temperature resistivity measurement
has  the same trends like the cyclotron mass, as shown in Figure~\ref{fig_quantum_oscillations_meff} \cite{Bristow2020}
Note that these values of the $A^{1/2}$ coefficient agree with those reported  in Ref.~\cite{Licciardello2019},
once adjusted for the correct value of $x$
based on the $T_s$ of each sample,
shown by the open triangle in Figure~\ref{fig_quantum_oscillations_meff}b.

To asses the nematic critical behaviour in FeSe$_{1-x}$S$_x$, it is worth emphasizing that the electronic correlations and the orbitally averaged cyclotron masses
do not show any divergence close to NEP ($x \sim 0.175 (5)$),
as shown in Figure~\ref{fig_quantum_oscillations_meff}c \cite{Coldea2019}.
Instead, the effective mass of FeSe$_{1-x}$S$_x$ reaches the
largest value  deep inside the nematic phase,
where the superconductivity is the strongest
and the low-energy spin-fluctuations are expected to be the largest \cite{Wiecki2018}.
The lack of divergent effective masses at NEP
points towards a finite
coupling of the electronic system with the underlying lattice
that can suppressed the critical nematic fluctuations,
except along certain directions in FeSe$_{1-x}$S$_x$\cite{Paul2017}.
Nematic susceptibility as a function of chemical pressure
suggest the possibility of having a nematic critical
point in FeSe$_{1-x}$S$_x$ \cite{Hosoi2016}.
However, at low temperatures there are no
divergent electronic correlation in any of the measured
quantities in the vicinity of the nematic end point,
suggesting an important role for the coupling of the electronic system with the lattice
in this system  \cite{Bristow2020,Coldea2019}.

 Signatures of quantum criticality caused by diverging spin fluctuations
were detected  in quantum oscillations in BaFe$_2$(As$_{1-x}$P$_{x}$)$_2$,
 by approaching the spin-density wave phase from the tetragonal phase.
 The cyclotron effective mass of the  outer electron bands increases
 from 1.8 to 3.5$m_e$ over a large compositional range ($x=0.4-1$)
in the tetragonal phase, as shown in Figure~\ref{fig_quantum_oscillations_meff}e \cite{Walmsley2013,Shishido2010}.
This enhancement of the effective mass in BaFe$_2$(As$_{1-x}$P$_{x}$)$_2$
correlates directly with the strong increase in the superconducting transition temperature.
The quantum oscillations frequencies originate from the lighter electron bands in BaFe$_2$(As$_{1-x}$P$_{x}$)$_2$
and their frequencies get smaller as the system evolves from the metallic tetragonal phase towards the spin-density wave phase.
These trends are similar to those expected for FeSe$_{1-x}$S$_x$.
It is worth emphasizing that for both systems only the effective mass is reported, not
the mass enhancement in relation to the band mass, due to the complexity involved in establishing the details of
the correct band structure for the mixed isoelectronic systems.
Figure~\ref{fig_quantum_oscillations_meff}
compares the effective masses for the two isoelectronic systems
and it suggests that the relevant interactions that enhance the effective masses in FeSe$_{1-x}$S$_x$ are the same
that enhance superconductivity.
These pairing interactions are strongest
 deep inside the nematic phase not at the nematic end point.
Their origin could be the spin fluctuations in both
systems and they are also likely to be
responsible for the linear resistivity observed inside the nematic phase for FeSe$_{1-x}$S$_x$ \cite{Bristow2020}
and for BaFe$_2$(As$_{1-x}$P$_{x}$)$_2$  for $x \sim 0.33$ \cite{Kasahara2014}.
 The superconductivity is strongly enhanced in the proximity of a magnetic critical point in BaFe$_2$(As$_{1-x}$P$_{x}$)$_2$,
as opposed to the small  abrupt drop in $T_c$ at the nematic end point FeSe$_{1-x}$S$_x$ (see Figure~\ref{fig_quantum_oscillations_meff}d).
This suggests that a strong nematoelastic effect suppresses the critical nematic fluctuations and the superconducting mechanism has a non-nematic origin in  FeSe$_{1-x}$S$_x$ \cite{Labat2017}.

\section{\bf The nematic susceptibility of FeSe$_{1-x}$S$_x$}

A direct measurement to test for the existence of an
intrinsic nematic electronic state is the determination of the nematic susceptibility,
that is the related to the  in-plane resistivity anisotropy
under a small amount of external strain \cite{Chu2012}.
These type of studies have established that the tetragonal-to-orthorhombic structural transition
in iron pnictides is  driven by the electronic instability of the system \cite{Chu2012}.
The Curie-Weiss behavior of nematic susceptibility near a nematic transition
is expected to display a generic mean-field behavior.
The nematic fluctuations of the nematic order parameter, which couple linearly to the orthorhombic
distortion via the nematoelastic coupling, are expected to be suppressed
but this may not be the case if the nematic fluctuations
are driven by the spin fluctuations \cite{Karahasanovic2016}.
The nematic susceptibility of Ba(Fe$_{1-x}$Co$_x$)$_2$As$_2$ follows a
Curie-Weiss dependence and the mean field nematic critical temperature closely tracks the actual structural transition temperature,
  being suppressed to zero at the optimal doping  \cite{Chu2012}.
 The divergence of the nematic susceptibility above $T_s$
indicates the tendency towards an electronic nematic phase transition and the Weiss temperature
indicates the strength of nematic fluctuations \cite{Chu2012}.
 At a critical nematic point, the nematic
 susceptibility should diverge at zero temperature (in proportion to $1/T$)
 and power law behaviours in temperature
 and composition are expected \cite{Palmstrom2019}.

Nematic susceptibility measurements for FeSe and FeSe$_{1-x}$S$_x$ \cite{Watson2015a,Tanatar2016,Hosoi2016}
indicate a large divergence above the $T_s$, similar to what was previously observed
in Ba(Fe$_{1-x}$Co$_x$)$_2$As$_2$, but in the absence of magnetic order \cite{Chu2012}.
Nematic susceptibility of FeSe
has an opposite sign to other pnictides \cite{Kuo2016},
but similar to other chalcogenides, like FeTe \cite{Zhang2012b},
as the resistance along the $a$ (AFM) direction is larger than that along $b$ axis (FM direction).
A sign-change in the in-plane anisotropy could be induced by
the different scattering rates by spin fluctuations  corresponding to different Fermi
velocities at the hot-spots for electron- and hole-doped pnictides \cite{Blomberg2013}.
In FeSe, despite the lack of long-range magnetic order,
 the anisotropy of the in-plane resistance below $T_s$ follows qualitatively
 a model assuming that the electrons are mainly scattered by magnetic fluctuations \cite{Tanatar2016,He2018}.
Elastoresistance measurements in FeSe$_{1-x}$S$_x$  superconductors found that the nematic transition temperature decreases with $x$,
 whereas the nematic fluctuations are strongly enhanced, similar to Ba(Fe$_{1-x}$Co$_x$)$_2$As$_2$.
 The  observation  of  strong nematic  fluctuations  is consistent with  the  presence  of  a  nematic quantum  critical  point,
 but this observation is insufficient to determine whether these fluctuations are driven by quantum criticality.
 Future studies to establish a suitable power law of the nematic susceptibility are needed \cite{Palmstrom2019}
  in order to identify whether this point represent a critical point in FeSe$_{1-x}$S$_x$.
  However, the  lack of  divergent  electronic correlations in quantum oscillations or enhanced superconductivity
 at the nematic end point suggest a strong suppression of critical fluctuations in FeSe$_{1-x}$S$_x$.

\section{\bf Magnetotransport behaviour of FeSe$_{1-x}$S$_x$ }

In multi-band systems with different carrier mobilities, the magnetoresistivity components $\rho_{xx}$ and $\rho_{xy}$
have complicated behaviour in magnetic field.
In the presence of a single dominant scattering time,
the magnetoresistance is expected to follow  Kohler's rule and a $B^2$ dependence \cite{Pippard1989}.
In the tetragonal phase of FeSe$_{1-x}$S$_x$, a quadratic dependence of the magnetoresistance is found up to 69~T both
at high temperature above $T_{\rm s}$ or at low temperature outside
the nematic phase boundaries for $x \geq 0.19$
\cite{Bristow2020}.
On the other hand, inside the nematic phase of FeSe$_{1-x}$S$_x$,
Kohler's rule is violated
and the magnetoresistance of FeSe$_{1-x}$S$_x$,
follows an unusual  $ B^{\sim 1.55}$ power law in high magnetic fields  \cite{Bristow2020}.
Furthermore, scaling to a modified Kohler's rule as a function of the Hall angle was found in
the vicinity of the nematic end point \cite{Huang2020}.
Another way to understand this complex behaviour is to
 separate different components of magnetoresistance, as suggested in Ref.~\cite{Licciardello2019MR}
 The coefficient of one of the extracted component has the same temperature
dependence as the resistivity slope in  34~T in FeSe (Fig.~\ref{fig_magnetotransport}),
once adjusted for the correct compositions as Refs.~\cite{Licciardello2019,Licciardello2019MR} uses  the nominal sulphur concentration.
Furthermore, other types
of magnetoresistivity scaling found for BaFe$_2$(As$_{1-x}$P$_{x}$)$_2$ \cite{Hayes2016},
are not found for FeSe$_{1-x}$S$_x$ \cite{Bristow2020,Licciardello2019MR},
and these effects are likely to occur for samples with higher impurity scattering \cite{Licciardello2019MR}.

\subsection{\bf Highly mobile small carriers in FeSe$_{1-x}$S$_x$
 beyond the two-band model}
\begin{figure*}[htbp]
	\centering
	\includegraphics[trim={0cm 0cm 0cm 0cm}, width=1\linewidth,clip=true]{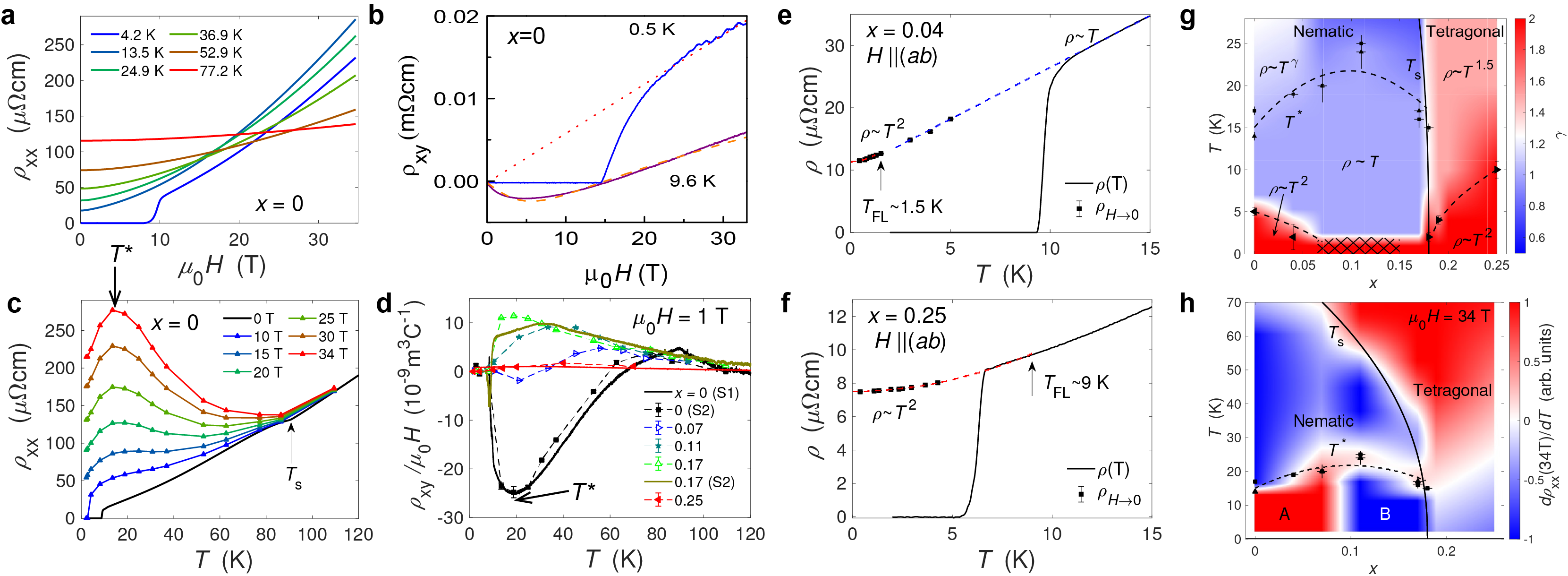}
	\caption{{\bf Magnetotransport behaviour of FeSe$_{1-x}$S$_x$.}
a) Longitudinal magnetoresistance, $\rho_{xx}$ and b) Hall component, $\rho_{xy}$
of FeSe at different fixed temperatures, after Refs.~\cite{Bristow2020,Watson2015c}.
The dotted and dashed lines are fits to two-band and three-band models, detailed in \cite{Watson2015c}.
c) The temperature dependence of magnetoresistance in fixed magnetic field indicating
significant effects  inside the nematic phase. Below $T*$
there is an unusual drop in magnetoresistance of FeSe 
suggesting potential changes in scattering and/or the electronic structure.
d) Hall effect coefficient in low magnetic fields
($B=\mu_0 H \leq 1$T), indicating a sign change and the dominance of different highly mobile carriers across the nematic phase.
The low-temperature linear resistivity for $x$=0.04 (e) and $x$=0.25 (f).
The solid lines are the zero-field resistivity data.
Solid circles represent the zero-field extrapolated values from high field longitudinal resistance measurements
when $B||(ab)$ plane \cite{Bristow2020}.
The dashed lines represent fits to  a Fermi-liquid behaviour found below $T_{FL}$,
as indicated by arrows.
(g) The low-temperature resistivity exponent below $T^*$, extrapolated from high magnetic fields,
		indicating the unusual transport behaviour of the nematic phase.
		(d) The colour plot of the slope of resistivity in 34~T between the nematic A and B phases.
		Solid squares represent $T_{s}$ and solid triangles $T_c$. $T^*$ indicated by stars
		represents the peak in magnetoresistance and the maximum in of the Hall coefficient $ R_H $.
		Solid lines indicate the nematic and superconducting phase boundaries and the dashed lines are guides to the eye.
		All the samples show clear quantum oscillations in the transverse magnetoresistance below 1.5~K ~\cite{Coldea2019},
consistent with the presence of the Fermi-liquid behaviour at low temperatures for all samples.
These data are adapted after Ref.~\cite{Bristow2020}.}
	\label{fig_magnetotransport}
\end{figure*}

 A compensated two-carrier model
 can describe the behaviour of the magnetoresistance and the Hall effect of FeSe$_{1-x}$S$_x$
in the tetragonal phase \cite{Watson2015b,Bristow2020}.
 For a compensated metal, the sign of the Hall coefficient depends on the difference
between the hole and electron carrier mobilities  \cite{Watson2015b}.
At high temperatures in the tetragonal phase, the Hall effect is linear and
the Hall coefficient, $R_{\rm H} = \rho_{xy}/ B$, is close to zero
as the hole and electron pockets have rather similar mobilities \cite{Watson2015b}.
Inside the nematic phase, the Hall coefficient for FeSe, extrapolated in the low-field limit (below 1~T), is negative at low temperatures with a minimum around 20~K \cite{Watson2015b,Bristow2020}.
These anomalies indicate that the magnetotransport behaviour of FeSe cannot be described using a two-band compensated model,
and an additional higher mobility charge carrier
is required (with carrier concentration of
0.7$\times 10^{20}$ cm$^{-3}$, a factor 5 smaller than the size of the largest band) \cite{Watson2015b,Huynh2016,Sun2016mob}.
This component would correspond to a Fermi surface of small volume
that could be linked to the inner electron band  with frequencies below 100~T
and a light effective mass ($\sim $ 2~$m_{\rm e}$) \cite{Terashima2014,Watson2015a,Coldea2019}.
With increasing $x$, the Hall coefficient at low temperatures is positive for  $x \sim 0.11$,
 suggesting the dominance of highly mobile hole carriers, as shown in Figure~\ref{fig_magnetotransport}d \cite{Bristow2020}.
This sign change occurs for the same composition
at which the highly mobile 3D hole pocket center is detected
in  ARPES studies at the $Z$-point around $x \sim 0.11$,
shown in Figure~\ref{Fig_ARPES} \cite{Watson2015c}.
High mobility carriers have been suggested
to dominate the magnetotransport behaviour across
the whole nematic phase of FeSe$_{1-x}$S$_x$ \cite{Sun2016mob,Ovchenkov2017},
Furthermore, FeSe under pressure shows similar trends to the chemical pressure effect
and the normal-state Hall resistivity changes sign
from negative to positive, showing dominant highly mobile hole carriers at high pressures \cite{Sun2017}.

The orbital order significantly affects band shifts for the electron bands
and it can generate very small pockets, as shown in Fig.~\ref{Fig_simulations_ARPES}.
The presence of the small number of highly mobile carrier was suggested to be
linked to the Dirac-like dispersions in the nematic phase on some sections of the electron pockets
 \cite{Tan2016,Huynh2016}.
Large orbital-dependent shifts of $\Delta_{M} \sim 70$~meV in FeSe thin films on SrTiO$_3$ have
been found to generate Dirac-like dispersion around the M point
\cite{Zhang2016}, but these shifts are much larger than those of  $\Delta_{M} \sim 50$~meV in bulk FeSe.
Magnetotransport cannot distinguish whether there are two tiny electron-like pockets or one small electron pocket
 (Fig.~\ref{Fig_simulations_ARPES}c for large $\phi_3$ values),
besides the almost compensated hole and electron bands in FeSe.
It is clear that a two-band picture containing an single electron and hole pocket
and assuming isotropic scattering fails  to describe magnetotransport behaviour of  FeSe \cite{Watson2015b}.

Significant changes in scattering could occur
for a elongated nematic Fermi surface of FeSe$_{1-x}$S$_x$
\cite{Watson2015a,Coldea2017}.
 Two scattering wave-vectors are detected by STM \cite{Hanaguri2018}
suggesting different scattering processes along certain directions
of an elongated ellipse \cite{Wang2019}.
A flower-shaped electron orbit would have a strongly varying
angular velocity \cite{Ong1991} and the scattering rate could vary strongly due to the changes of the orbital character on various sections
induced by spin fluctuations \cite{Tanatar2016,Luo2017}.
Hall effect in iron-based superconductors is also affected
by the spin fluctuations that induce mixing of the electron and hole currents \cite{Fanfarillo2012}.
All these effects could lead to highly anisotropic scattering rates
in FeSe$_{1-x}$S$_x$  that are suppressed with $x$ substitution.
Indeed, in the tetragonal phase a single scattering process dominates the magnetotransport, as Kohler's rule is obeyed \cite{Bristow2020}.
Further theoretical work is needed to understand
transport and magnetotransport data of FeSe$_{1-x}$S$_x$.
Future models should account for anisotropic scattering and scattering of quasiparticles from the domain walls,
when the nematic domain size (determined by the  quenched disorder)
 is smaller than the normal state mean-free path \cite{Fradkin2010}.

FeSe$_{1-x}$S$_x$  displays deviation from the Fermi liquid theory,
expected for conventional metals,
 that affect the temperature and field dependencies of electron transport.
The magnetoresistance of FeSe$_{1-x}$S$_x$
increases significantly once a system enters the nematic state and
shows an unusual temperature dependence that varies strongly
with $x$, as shown in Figure~\ref{fig_magnetotransport}a and b for FeSe.
The temperature dependence of the resistivity slope in  34~T in FeSe
changes sign at a characteristic temperature, $T^*$ below 14~K,
and the Hall coefficient $R_H$  display a negative maximum, as shown in Fig.~\ref{fig_magnetotransport}b and d.
Interestingly,  $T^*$  seems to be the characteristic scale for
low-energy spin fluctuations in FeSe$_{1-x}$S$_x$ \cite{Wiecki2018,Shi2018}.
Magnetostriction measurements in magnetic field for FeSe showed that the lattice distortion
continues to increase down to $T_c$,  different from BaFe$_2$As$_2$, where there is a intimate connection between the magnetic order and structural
distortion \cite{He2018}.
With sulphur substitution, $T^*$  shifts to a slightly higher temperature
of $\sim 20$~K, and eventually disappears in the tetragonal phase, as the
low-energy spin fluctuations are completely suppressed \cite{Wiecki2018,Grinenko2019}.
Changes in magnetotransport and in the resistivity slope
occur from $x\sim 0.11$ (Fig.~\ref{fig_magnetotransport}h)
in the presence of the additional highly mobile 3D band,
labelled as the nematic B phase  \cite{Bristow2020}.
It is clear that magnetic field could affect scattering inside the nematic phase
that could be still dominated by spin fluctuations
and it can spin polarize the multi-band small  Fermi-surface
of FeSe$_{1-x}$S$_x$. Further theoretical work will be required to
explain the observed effects in magnetic fields
and experimental studies in single domains crystals are needed to address
the extrinsic scattering at the nematic domain boundaries.

\subsection{\bf Anomalous transport behaviour}
 Linear resistivity at low temperatures is usually found near an antiferromagnetic critical point,
 such as in BaFe$_2$(As$_{1-x}$P$_x$) \cite{Kasahara2010} and reflect  scattering induced by critical
spin-fluctuations  \cite{Rosch1999}.
 FeSe$_{1-x}$S$_x$ has a low temperature region with a linear resistivity
 across the whole nematic phase below $T^*$ (using extrapolated in-plane high magnetic field data), as shown in Figure~\ref{fig_magnetotransport}e and g.
Low energy spin-fluctuations are present inside the nematic state in FeSe$_{1-x}$S$_x$
\cite{Wiecki2018,Shi2018,Grinenko2019},
and $\mu$SR studies place FeSe
near an itinerant antiferromagnetic quantum critical point at very low temperatures \cite{Grinenko2018}.
This region with linear T resistivity below $T^*$
occurs over a limited temperature regime and the Fermi-liquid behaviour recovers below $T_{FL}$
and all compositions show quantum oscillations (Fig.~\ref{fig_magnetotransport}e-g) \cite{Coldea2019}.
Deviations from  Fermi-liquid behaviour were also reported for FeSe \cite{Kasahara2010}
and linear T resistivity was detected in 35~T  for $x_{nom} \sim 0.16$ with $T_s \sim 50$~K,
which corresponds to $x \sim 0.13$ inside the nematic phase ~\cite{Licciardello2019,Shibauchi2020}.
Thus, all existing experimental transport data for FeSe$_{1-x}$S$_x$ in high magnetic fields up to 45~T
suggest that  the low-temperature linear resistivity occurs inside the nematic phase, rather than at the nematic end point,
and, as in the case of the electronic correlations
and $T_c$,
it is likely a manifestation of the spin fluctuation scattering mechanism inside the nematic phase FeSe$_{1-x}$S$_x$.

In the tetragonal phase of FeSe$_{1-x}$S$_x$, the resistivity exponent seem to vary with temperature \cite{Bristow2020}
and a resistivity with $T^{3/2}$ dependence is found over a large temperature range up to 120~K,
in agreement with studies under pressure \cite{Reiss2020}.
Theoretical models suggest that the exact temperature exponent of resistivity,
in vicinity of nematic critical points is highly dependent
on the presence of {\it cold spots} on different Fermi surfaces, due to the symmetry
of the nematic order parameter \cite{Wang2019,DellAnna2007,Maslov2011} or
due to the scattering from acoustic phonons \cite{Carvalho2019} near the nematic end point.
Near a Pomeranchuk instability the transport decay rate is linear in temperature everywhere on the Fermi surface except
at cold spots on the Brillouin zone diagonal,
 leading to a resistivity proportional to $T^{3/2}$  for a clean 2D system
or to a linear $T$ dependence in the presence of impurities \cite{DellAnna2007}.
Furthermore, the scale at which the crossover to Fermi liquid behavior occurs at low temperatures
 depends on the strength of the coupling to the lattice \cite{Paul2017},
 responsible for the lack of divergent critical fluctuation at the nematic end point in
 FeSe$_{1-x}$S$_x$ \cite{Coldea2019,Bristow2020,Reiss2020}.

\section{\bf Superconductivity of FeSe$_{1-x}$S$_x$}
The normal nematic electronic phase and the
anomalous scattering of FeSe$_{1-x}$S$_x$ affects significantly
its superconducting state.
The gap structure of FeSe is two-fold symmetric,
reaching small values on the major axis of the elliptical hole pocket
and it is changing its sign between the hole and
the small electron pocket \cite{Sprau2016,Xu2016}.
While nematicity is an
 intrinsic property of the bulk FeSe$_{1-x}$S$_x$,
 nematic fluctuations may not be the primary force driving the superconducting pairing \cite{Acharya2020},
 despite the fact that the relative orthorhombic distortion is
 reduced as the superconductivity increases in  FeSe$_{1-x}$S$_x$  \cite{Wang2016Meingast}.
Instead, the low-energy spin-fluctuations are likely to provide the pairing channel in FeSe \cite{Imai2009}
and this can manifest via nesting of $d_{yz}$ sections of the hole and
electron bands; the $d_{xy}$ portions do not participate in pairing
due to the orbital selective strong correlation effects
\cite{Kreisel2017,Sprau2016}.
In this scenario, a maximum gap on the Fermi surface sections with $d_{yz}$ character,
and a small gap on sections with $d_{xz}$ or $d_{xy}$ character would occur, similar to experiments \cite{Sprau2016,Rhodes2018}.
Most of the thermodynamic and thermal conductivity studies of bulk FeSe in the superconducting phase have been modelled by accounting
for two different nodeless superconducting gaps
\cite{Lin2011,Bourgeois-Hope2016}.
The presence of nodes in the superconducting gap of FeSe has also been suggested by other studies \cite{Song2011,Moore2015,Kasahara2014}.
The multi-gap superconductivity is preserved as a function of chemical pressure in FeSe$_{1-x}$S$_x$ \cite{Xu2016,Abdel-Hafiez2015},
and tunnelling experiments
found that the vortex core anisotropy is strongly suppressed once Fermi surface becomes isotropic \cite{Moore2015}.
High-resolution thermal expansion
 showed a lack of coupling between the orthorhombic distortion and superconductivity in FeSe \cite{Bohmer2013},
 whereas with increasing substitution towards $x \sim 0.15 $ the effect seems to be the opposite \cite{Wang2016Meingast}.
The jump in specific heat ($\Delta C_{el}$ /$\gamma_n$ $T_c$) for different $x$
  varies slightly around 2, which is above the weak
  coupling limit of the BCS theory believed to be caused by the multi-band effects
  \cite{Abdel-Hafiez2015}.
 For isotropic isoelectronic iron-based superconductors,  the height of pnictogen
acts as a switch between high-$T_c$ nodeless and low-$T_c$ nodal pairings \cite{Kuroki2009}.
 FeS, like other end member compounds, displays weak correlations and nodal superconductivity,
 similar to other systems like LaFePO and LiFeP  \cite{Fletcher2009,Hashimoto2012,Xing2016,Yang2016},
as the chalcogen position is closer to the iron planes
compared to their isoelectronic sister-compounds, like LiFeAs.
On the other  hand for FeSe,
there has been suggestions both of nodal and nodeless superconductivity \cite{Hardy2019,Kasahara2014,Bourgeois2016,Jiao2017}.
  Abrupt changes in the superconductivity occur
at the nematic end point, potentially stabilized
by different pairing channels inside and outside the nematic phase
\cite{Sato2018,Hanaguri2018}.
There is no superconductivity enhancement  at the nematic end point in FeSe$_{1-x}$S$_x$,
suggesting the presence of a non-nematic pairing mechanism and/or
the lack of divergent critical fluctuations, similar to the behaviour of
the quasiparticle effective masses  \cite{Coldea2019,Reiss2020}.
The coupling to the relevant lattice strain restrict the
 critical behaviour only along certain high symmetry directions
 and this can affect the nematic critical fluctuations and
 do not enhance superconductivity \cite{Labat2017,Paul2017}.

\subsection{\bf BCS-BEC crossover of the multiband FeSe$_{1-x}$S$_x$}
 FeSe$_{1-x}$S$_x$ are multi-band systems with relatively small Fermi energies at low temperatures.
There has been a lot of interest to asses
whether these systems are candidates in the crossover regime between the BCS to the BEC state,
expected for  $\Delta_{SC}/E_{\rm F}  \leq 1 $  \cite{Randeria2014,Chen2005}.
These effects have been suggested to occur in Fe$_{1+y}$Se$_x$Te$_{1-x}$, as the
hole band as the $\Gamma$ point is tuned at the Fermi level by Fe deficiency
and  $\Delta_{SC}/E_{\rm F} $ varies 0.16 to 0.50 \cite{Rinott2017}.
This ratio is also relevant for assessing the possibility of stabilization of a FFLO state
in FeSe \cite{Kasahara2020} and a good knowledge of the value of
the superconducting gap and the Fermi energy of the multi-band and highly warped Fermi surfaces is needed.
The amplitudes of the highly anisotropic superconducting gaps of FeSe around the hole pocket
vary between $\Delta_{SC}  \sim $2.5 (or 2.3)  ($\Gamma$ point) from STM
to 1.5~meV to 3~meV from laser ARPES \cite{Hashimoto2018}.
For the electron pocket (at the M point)
the values of the gap vary between 3.5 or 1.5~meV \cite{Hanaguri2018,Sprau2016} and
another potential small gap of 0.39~meV was invoked from specific heat data \cite{Sun2017}
(Fig.~\ref{Fig_ARPES_paramaters}f).
The top of the hole band (associated to the Fermi energy $E_{\rm F}$) is $k_z$ dependent
having a value of  16~meV at $\Gamma$ point and  25~meV at the $Z$ point
(Fig.~\ref{Fig_ARPES_paramaters}d) \cite{Watson2015a,Coldea2017},
whereas laser ARPES  reports values of 6.7-10~meV  \cite{Hashimoto2018,Liu2018}.
Based on these values, $\Delta_{SC}/E_{\rm F}  \sim 0.1-0.15 $
for the hole band at the $\Gamma$ point in FeSe.
This ratio will decrease using the parameters at Z point and the correct values of $E_{\rm F}$ need
to take into account the strong $k_z$ dependence of the cylindrical Fermi surface
and the mass anisotropy for each pocket.
As a function of $x$,  the gap associated with the hole
band remains relatively constant inside the nematic phase,
but it is getting smaller towards 1.5~meV in the tetragonal phase
(see Fig.~\ref{Fig_ARPES_paramaters}f) \cite{Hanaguri2018}.
The top of hole band and the $\Gamma$
is somewhat reduced towards 5~meV for $x\sim 0.18$
 but  increases slightly for the Z point at 26~meV;
 the bottom of the inner electron band
is around $\sim  15$~meV (see Fig.~\ref{Fig_ARPES_paramaters}f).
The variation of these parameters will
affect the estimates of $\Delta_{SC}/E_{\rm F}  $
and one needs to consider the multiple bands and gaps of FeSe$_{1-x}$S$_x$
together with the $k_{z}$ dependence of the Fermi surface and
the superconducting gap \cite{Kushnirenko2018,Rhodes2019}.
Furthermore,  the Fermi velocities increase with $x$
 pushing the system away from the BCS-BEC crossover regime,
 as not all the bands satisfy the crossover condition.

Another way to assess the proximity to the crossover is to check whether
the size of the Cooper pair, given by the coherence length $\xi_{ab}$,  is smaller than the mean inter-particle spacing $1/k_F$
and $\xi k_F \ll 1 $\cite{Chen2005,Kreisel2020}. Using the in-plane coherence length for FeSe
of $\xi $ = 4.6-5.7~nm \cite{Terashima2014,Bristow2020Hc2}
and the values of  $k_F \sim 0.038 -0.15$ for the hole bands (Fig.~\ref{Fig_ARPES_paramaters}),
it suggests that $\xi k_F  \sim 1.75-8.55$ is large and the Cooper pairs are quite extended
suggesting that the superconductivity of FeSe need to be
understood considering its multi-band effects.
Further aspects of the pairing mechanism of FeSe and other iron-chalcogenides
are discussed in detail in recent reviews \cite{Kreisel2020,Shibauchi2020}.

\section{\bf Conclusion}
FeSe$_{1-x}$S$_x$  has opened an new area of exploration of the electronic nematic state
 and its role in the stabilization of the unconventional superconductivity.
 These systems are multi-band systems which are highly sensitive
to orbitally-dependent electronic interactions  that affect the evolution of the electronic structure
with sulphur substitution.
Fermi surface of FeSe$_{1-x}$S$_x$ are mainly quasi-two dimensional warped cylinders
but an additional 3D hole pocket is present with increasing sulphur concentration from $x\sim 0.11$.
The Fermi energies have a broad range,  that generally increases with $x$ substitution,
but they are smaller for the inner electron and hole pockets, making them prone to electronic instabilities.
The development of nematic electronic phase with strong anisotropic electronic structure
influences the scattering and leads to unusual magnetoresistance inside the nematic phase.
Linear resistivity and anomalous magnetotransport
is detected inside the nematic phase and is likely to reflect
the role played by the spin fluctuations in this regime.
FeSe$_{1-x}$S$_x$ show no signatures
of enhanced $T_c$ and divergent electronic correlations
at the nematic end point, which are likely to be quenched by the finite coupling with the lattice.
This coupling could also be the origin of the non-Fermi liquid behaviour outside the nematic phase.
The superconductivity of FeSe$_{1-x}$S$_x$ has a small enhancement  inside the nematic phase
and a somehow abrupt change at the nematic end point.
This behaviour is different from the isoelectronic family BaFe$_2$(As$_{1-x}$P$_x$)$_2$
where quantum critical fluctuations enhance both superconductivity and effective masses of the quasiparticles
on approaching the spin-density phase and linear resistivity is found at the magnetic critical point.
The study of FeSe$_{1-x}$S$_x$ compared with other
isoelectronic iron-based superconductors  emphasis the important role played
by the magnetic rather than nematic fluctuations for
enhancing superconductivity in iron-based superconductors.

\section{\bf Acknowledgements}

I am very grateful to numerous collaborators for their important scientific contribution dedicated to the understanding
of the electronic structure of FeSe$_{1-x}$S$_x$.
I would like to thank warmly to  Amir Haghighirad, Shiv Singh, Thomas Wolf and Shigeru Kasahara for the growth of high quality crystals;
Timur Kim, Matthew Watson, Pascal Reiss and Kylie MacFarquharson for for their contributions to ARPES studies; 
Pascal Reiss,  Matthew Bristow, Zachary Zajicek, Samuel Blake, Mara Bruma, David Graf, Alix McCollam,William Knafo
for their contributions to high magnetic field studies; Joe Prentice, Roemer Hinlopen, Oliver Humphries, Oliver Squire, Gladys Lee, James Bate,
Andrew Smith, Tom Nicholas, Abhinav Naga, Jan Memedovic, Joshua Bell, Michele Ghini,
 Nathaniel Davies for their individual contributions to experimental and computational aspects
 that have enhancee our research.
I am grateful to Rafael Fernandes for sharing his computer code and Andreas Kreisel, Peter Hirschfeld, M. Watson, Roser Valenti,
Karim Zantout and Zachary Zajicek for useful comments on this manuscript.
I acknowledge the financial support provided by EPSRC  (EP/I004475/1,  EP/I017836/1),
Oxford Centre for Applied Superconductivity
and Diamond  Light  Source  for  experimental access  to  the I05 Beamline.
I am also grateful for an EPSRC Career Acceleration Fellowship (EP/I004475/1).

\bibliography{FeSeS_bib_may2020}

\end{document}